%% file: workfile-preprint.tex
\documentclass[10pt]{article} 
\usepackage{amsmath}
\usepackage{amssymb}
\usepackage{cite}
\usepackage{indentfirst}
\usepackage{verbatim} 
\usepackage{graphicx}

\def\de{\delta}
\def\ds{\displaystyle}
\def\ni{{\noindent}}
\def\di{{\partial}}
\def\ie{{\it i.e.~}}

\def\xp{{{ x}^{\prime}}}

\def\zp{{{ z}^{\prime}}}
\def\zpp{{{ z}^{\prime\prime}}}
\def\zppp{{{ z}^{\prime\prime\prime}}}

\def\bp{{\bf p}}

\def\h!{\!\!\!}
\def\f!{\!\!\!\!\!\!}

\def\e{\textrm{e}}
\def\tx{\textrm}
\def\cL{{\cal L}}      
\def\cH{{\cal H}}      
\def\hH{\hat{H}}       
\def\hP{\hat{P}}
\def\x{{\bf x}}        
\def\r{{\bf r}}
\def\v{{\bf v}}
\def\p{{\bf p}}
\def\q{{\bf q}}
\def\cY{{\cal Y}}           
\def\cC{{\cal C}}           
\def\cQ{{\cal Q}}
\def\b{\beta}
\def\i{\textrm{i}}          
\def\E1{\textrm{E}_{1}}     
\def\Li2{\textrm{Li}_{2}}   
\def\hR{\hat{R}} 
\def\hv{\hat{v}}
\def\vr{\varrho}            
\def\F!{\;\;\;\;\;\;\,\,}
\def\H!{\;\;\;}
\def\bra{{\langle}}
\def\ket{{\rangle}}

\newcommand{\bft}[1]{\textbf{#1}}

\newcommand{\re}[1]{(\ref{#1})}
\newcommand{\nn}{\nonumber}
\begin{document}

\title{Inter-particle potentials in a scalar QFT with a Higgs-like mediating field}
\author{Alexander Chigodaev and Jurij W. Darewych \\ Department of Physics and Astronomy 
\\York University, Toronto, ON, M3J 1P3}
\maketitle
\begin{abstract}
  We study the inter-particle potentials for few-particle systems in a scalar theory with a non-linear mediating field of the Higgs type. We use the variational method, in a reformulated Hamiltonian formalism of QFT, to derive relativistic three and four particle wave equations for stationary states of these systems. We show that the cubic and quartic non-linear terms modify the attractive Yukawa potentials but do not change the attractive nature of the interaction if the mediating fields are massive.
\end{abstract}
\input{introduction}

\input{reformulation}
\input{quantization}
\input{two}
\input{three} 
\input{four}
\input{conclusion}
\input{appendixA}
\input{appendixB}

\input{appendixC}
\bibliographystyle{ieeetr}
\bibliography{references}
\end{document}

%% file: introduction.tex
\section*{Introduction}
Inter-particle interactions in systems described by QFTs, such as QED or the Yukawa model and its variants (like the boson exchange model in nuclear theory), are specified by the interaction term (or terms) of the underlying Lagrangian density. For theories where the mediating fields are linear ({\sl e.g.} the Electromagnetic field for QED or the boson fields in nuclear theory), the relevant interaction terms of the Lagrangian (density) are of the generic form $\Psi{^\dagger} A \, \Psi$, where $\Psi$ is the particle field (charged fermions in QED; nucleons in nuclear theory) and $A$ describes the mediating field 
(photons in QED and mesons in nuclear theory).  Few particle wave equations that stem from theories with such interaction terms in the Lagrangian density are, in the non relativistic limit,  Schr\"odinger equations with inter-particle potentials given by $\ds \pm \, \alpha \frac{e^{- \mu r}}{r}$, where $\alpha$ is the coupling constant, $\mu$ is the mass of the mediating-field quantum and $r$ is the inter-particle distance. However, there are theories in which the mediating fields are non-linear, such as, for example, in the Higgs model, QCD  and  other models. In these theories, the interaction part of the Lagrangian density includes terms that are typically of degree 3 and 4 in the mediating field $A$. The analytic derivation of the inter-particle potentials in such cases becomes a more difficult and generally not achievable task.

In this paper, we shall address this problem for a scalar field theory with a non-linear mediating field, specified by the Lagrangian density (units: $\hbar=c=1$):
\begin{equation} 
  \cL=\di^\nu\phi^{\ast} \, \di_{\nu}\phi - m^2 \, \phi^{\ast} \, \phi - g \, \chi \, \phi^{\ast} \phi - \lambda \left(\phi^{\ast} \phi\right)^2 + \frac{1}{2}\left(\di^{\nu}\chi \, \di_{\nu}\chi - \mu^2 \, \chi^2\right) - \frac{1}{3} \, \eta \, \chi^3 - \frac{1}{4} \, \sigma \, \chi^4.
  \label{LAGRANGIAN}
\end{equation}
The parameters $m$ and $\mu$ represent the bare masses of the scalar and mediating field quantum respectively, $g, \eta$ are coupling constants with dimensions of mass, while $\sigma, \lambda > 0$ are dimensionless coupling constants. Note that the $\chi$ field is of the form of the Higgs field of the Standard Model. Moreover, the non-linear terms in the field $\chi$ are of a form that mimics those of the gluon field in the QCD Lagrangian if $\mu = 0$. The $\lambda$ term in the Lagrangian density is required to keep the classical ground energy bounded from below. It leads to a repulsive contact (delta function) inter-particle interaction for few-particle systems~\cite{Beg:1984yh, Darewych:1997uc}. Henceforth, we set $\lambda = 0$ since it has negligible effect on what follows in this work

The aim in this paper is to derive relativistic wave equations for few-particle states of this model, with an emphasis on the study of the inter-particle interaction potentials. We shall use a reformulated Hamiltonian formalism, together with the variational method, to accomplish this task. This approach has been used effectively in the study of few-boson states in the scalar Yukawa theory~\cite{Ding:1999ed, EmamiRazavi:2006yx} and relativistic few-fermion bound states in relativistic quantum mechanics~\cite{Grandy1991} and QED~\cite{Terekidi:2003gp, Barham:2007vd}.

%% file: reformulation.tex
\section*{Reformulation}
We introduce a reformulation of the Lagrangian density with the intention of expressing the effect of the mediating field $\chi$ through its propagator. The reformulation corresponds to a partial solution of the equations of motion. The equations of motion corresponding to the Lagrangian \re{LAGRANGIAN}, derived from the action principle $\delta\, \ds\int dx\, {\cal L} = 0$, are 
\begin{gather}
  \left(\di^{2} + m^2 \right) \, \phi(x) = \, -g\, \phi(x) \, \chi(x),  
  \label{EQ:PHI}\\
  \left(\di^{2} + \mu^2 \right) \, \chi(x) = \, -g \, \phi^{*}(x) \, \phi(x) - \eta \, \chi^2(x) - \sigma \, \chi^3(x). 
  \label{EQ:CHI}
\end{gather}
Equation (\ref{EQ:CHI}) has the integral representation (\ie ``formal solution'')
\begin{equation} 
  \chi(x) = \chi_0(x) + \int d\xp \,D(x-\xp)\, \rho(\xp),
  \label{EQ:CHI_FORMAL_SOL}
\end{equation}
where $x = (t, \x)$, $dx = dt \, d\x $, and 
\begin{equation}
  \rho(x) =  - g \phi^*(x) \phi (x) - \eta \, \chi^2(x) - \sigma \chi^3(x)
\end{equation}
is the ``source term'' of the inhomogeneous equation. The function $\chi_0 (x)$ satisfies the homogeneous (or free field) equation with the right hand side of \re{EQ:CHI} equal to zero, while $D(x-\xp)$ is a covariant Green function (or the Feynman propagator) for the mediating field $\chi$, such that
\begin{equation}
  \left ( \di^2  + \mu^2 \right ) D(x-\xp) = \delta^{(4)}(x-\xp).
\end{equation}
Recall that the Green function can be expressed as
\begin{equation}
  D(x-\xp)=D(x,\xp)=\lim_{\epsilon \rightarrow 0} \int \frac{dk}{(2\pi)^{4}} \frac{1}{\mu^{2}-k^{2}+i\epsilon} \textrm{e}^{-i k(x-\xp)}.
  \label{EQ:GREEN} 
\end{equation}

We shall not be concerned with processes that involve free quanta of the mediating field $\chi_0$, and so henceforth we shall leave it out. Then, substituting (\ref{EQ:CHI_FORMAL_SOL}) into (\ref{EQ:PHI}), omitting terms containing $\chi_0$, we obtain the equation  
\begin{multline}
    \left(\di^2 + m^2\right) \phi(x) = g^2 \phi(x) \int d\xp \phi^*(\xp) \phi (\xp) D(x,\xp) \\ + g \eta \, \phi(x)  \int d\xp \, \chi^2(\xp) \, D(x,\xp) + g \sigma \, \phi(x) \int d\xp \chi^3(\xp) \, D(x,\xp). 
  \label{EQ:MOD_CHI}
\end{multline}
Equation \re{EQ:MOD_CHI} is derivable from the Lagrangian density
\begin{multline}
  \cL = \di_{\mu} \phi^{\ast}(x) \, \di^{\mu} \phi (x) - m^2 \, \phi^{\ast}(x) \phi (x) + \frac{1}{2}\,  g^2 \int d\xp \phi^{\ast}(x) \phi(x) D(x,\xp) \phi^{\ast}(\xp) \phi(\xp) \\ +  g\eta \, \phi^{\ast}(x) \phi(x) \int d\xp D(x,\xp)\chi^2(\xp) +   g \sigma \, \phi^{\ast}(x) \phi(x) \int d\xp D(x,\xp)\chi^3(\xp)
  \label{MOD_LAGRANGIAN}
\end{multline}
provided that the Green function is symmetric, $\ie D(x-\xp) = D(\xp-x)$.

For the case of a linear mediating field, \ie when $\eta = \sigma =0$, the reformulated Lagrangian density (\ref{MOD_LAGRANGIAN}) gives field equations that involve the particle fields only; the mediating field $\chi$ appears only through the propagator $D(x,\xp)$. The reformulated Lagrangian density (\ref{MOD_LAGRANGIAN}), with $\eta = \sigma =0$, is convenient for the study of few-boson relativistic bound states in the scalar Yukawa theory as will be pointed out below and as was shown in references~\cite{Ding:1999ed,EmamiRazavi:2006yx}. 

However, for the present case of a non-linear mediating field, \ie $\eta, \sigma \neq 0$, it is not possible to obtain a closed-form solution of (\ref{EQ:CHI_FORMAL_SOL}) for the field $\chi$ in terms of $D(x,x')$ and $\phi$, thus we need to resort to approximation schemes. An obvious one is an iterative sequence, based on \re{EQ:CHI_FORMAL_SOL}. The first order approximation corresponds to substituting the formal solution (\ref{EQ:CHI_FORMAL_SOL}), with $\rho = - g\, \phi^{\ast} \phi$ (with $\chi_0$ left out), into the Lagrangian density (\ref{MOD_LAGRANGIAN}). This yields the first-order iterative approximate expression
\begin{gather}
  \cL = \di_{\mu} \phi^{\ast}(x) \, \di^{\mu} \phi (x) - m^2 \, \phi^{\ast}(x) \phi (x) + \frac{1}{2}\,  g^2 \int d\xp \phi^{\ast}(x) \phi(x) D(x,\xp) \phi^{\ast}(\xp) \phi(\xp) \nn
  \\ + \;  g \, \eta \, \phi^{\ast}(x) \phi(x) \int d\xp \, d\zp \, d\zpp  \; D(x,\xp) \, D(\xp, \zp) \, D(\xp, \zpp) \; \rho(\zp) \, \rho(\zpp) \nn 
  \\ + \; g \, \sigma \, \phi^{\ast}(x) \phi(x) \int d\xp \, d\zp \, d\zpp \, d\zppp \; D(x,\xp) \, D(\xp, \zp) \, D(\xp, \zpp) \, D(\xp, \zppp) \, \rho(\zp) \, \rho(\zpp) \, \rho(\zppp).  
  \label{MOD_II_LAGRANGIAN}
\end{gather}
The interaction terms in the Lagrangian density (\ref{MOD_II_LAGRANGIAN}) do not contain the mediating field $\chi$ explicitly but rather implicitly through the mediating field  propagators. The advantage of reformulating the Lagrangian density to the present form will be evident when we implement quantization. As we shall see, this will enable us to use simple few-particle trial states free of the $\chi$ field quanta while still probing the effects of all interaction terms, including the non-linear terms (\ie those with $\eta, \sigma \neq 0$).  

The Hamiltonian and momentum densities corresponding to (\ref{MOD_II_LAGRANGIAN}) are obtained from the energy-momentum tensor in the usual way:
\begin{equation}
  T^{\mu\nu} = \frac{\di \cL}{\di (\di_{\mu} \phi_i)}\, \di^{\nu}\phi_i - g^{\mu\nu} \,\cL
\end{equation}
where the index $i = 1,2$ stands for the fields $\phi_1 = \phi$ and $\phi_2 = \phi^{\ast}$. The $T^{00}$ component of the energy-momentum tensor is the reformulated Hamiltonian density:
\begin{equation}
  \cH = {\dot \phi} \, \Pi_\phi + {\dot \phi}^{\ast} \, \Pi_{\phi^{\ast}}  - \cL =  \cH_{\phi} + \cH_{I_{1}} + \cH_{I_{2}} + \cH_{I_{3}},  
    \label{MOD_HAMILTONIAN}
\end{equation}
where
\begin{gather}
  \cH_{\phi} = \; \Pi_{\phi^{\ast}} \, \Pi_\phi + (\nabla {\phi^{\ast}}) \cdot (\nabla \phi) + m^2 \, \phi^{\ast} \phi,  \\
  \cH_{I_{1}} = - \, \frac{1}{2}\,  g^2 \phi^{\ast}(x) \phi(x)\int d\xp   \phi^*(\xp) \phi(\xp)D(x-\xp),   \\
  \cH_{I_{2}} = - g \, \eta \, \phi^{\ast}(x) \phi(x) \int d\xp \, d\zp \, d\zpp  \; D(x,\xp) \, D(\xp, \zp) \, D(\xp, \zpp) \; \rho(\zp) \, \rho(\zpp), 
  \label{MOD_HAM_C} \\
  \cH_{I_{3}} = - g \, \sigma \, \phi^{\ast}(x) \phi(x) \int d\xp \, d\zp \, d\zpp \, d\zppp \; D(x,\xp) \, D(\xp, \zp) \, D(\xp, \zpp) \, D(\xp, \zppp) \, \rho(\zp) \, \rho(\zpp) \, \rho(\zppp)
  \label{MOD_HAM_Q}.
\end{gather}
\noindent The conjugate momenta of the fields are defined in the usual way, $\ie \Pi_\phi = \ds\frac{\di \cal L}{\di {\dot \phi}} =  {\dot \phi}^{\ast}$ and $\Pi_{\phi^{\ast}} = \ds\frac{\di \cal L}{\di {\dot \phi}^{\ast}} = {\dot \phi}$.  Observe that every term of the reformulated Hamiltonian density has dimensions of $M^4$ (as it should). Notice that the Hamiltonian density is non-local, \ie it depends on more than one space-time variable. Equation (\ref{MOD_HAMILTONIAN}) is the Hamiltonian density that we shall use in the present work. 

The $T^{0i}$ components of the energy-momentum tensor are the momentum density components
\begin{equation}
  {\cal P}^i = - \Pi_\phi \, \di_i \, \phi - \Pi_{\phi^{\ast}} \, \di_i \, \phi^{\ast}.
  \label{MOMENTUM}
\end{equation}
Note that the interaction terms do not contribute to the momentum density above. We shall make use of this expression when the total momentum of the few-particle systems will be discussed. 

%% file: quantization.tex
\section*{Formalism and Quantization}
We will work in the Hamiltonian formalism of QFT where the basic equation to be solved is the 4-momentum eigenvalue equation
\begin{equation}
  \hP^{\nu} \,|\Psi \ket = Q^{\nu}\, |\Psi\ket,
  \label{EQ:EV}
\end{equation}
where $\hP^{\nu}$ is the energy-momentum operator of the quantized theory which follows from equations (\ref{MOD_HAMILTONIAN}) (as specified below) and (\ref{MOMENTUM}), $Q^{\nu} = (E, {\bf Q})$ is the energy-momentum eigenvalue and $|\Psi\ket$ are the corresponding eigenfunctions. It is not possible to obtain exact solutions for the $\nu = 0$ component of equation (\ref{EQ:EV}). Consequently, if one wished to solve for the energy of a system in question, then approximation methods must be used. For $\nu = 1,2,3$ the solutions of (\ref{EQ:EV}) are free-field-like as we point out below.

Variational approximations to the $\nu = 0$ component of the eigenvalue equation (\ref{EQ:EV}), \ie the Hamiltonian component, require the evaluation of
\begin{equation}
  \delta \bra \Psi_{t}|{\hat H} - E|\Psi_{t} \ket =  0,
  \label{EQ:VAR_PRIN}
\end{equation}
where $|\Psi_{t} \ket$ are suitably chosen trial states that contain adjustable features (functions, parameters). The applicability, accuracy and usefulness of a variational approximation depend on the choice of the trial states. They must be appropriately chosen for every case being studied. 

We now proceed to the canonical quantization of the theory with the Hamiltonian density (\ref{MOD_HAMILTONIAN}). First, we expand the fields in terms of its Fourier modes:
\begin{align}
  \phi(x) = & \int d\p\ds\left[(2\pi)^{3} \, 2 \omega_{\p}\right]^{-\frac{1}{2}} \big(a\left(\p\right)\e^{-i p \cdot x}+b^{\dagger}\left(\p\right)\e^{i p \cdot x}\big), \\
  \phi^{\ast}(x) = & \int d\p\ds\left[(2\pi)^{3} \, 2 \omega_{\p}\right]^{-\frac{1}{2}}\big(a^{\dagger}\left(\p\right)\e^{i p \cdot x}+b\left(\p\right)\e^{-i p \cdot x}\big),
\end{align}
where $\omega^2_{\p} = m^2 + \p^2$. Note that since the mediating field $\chi$ does not appear in the Hamiltonian explicitly, we need not be concerned with its Fourier mode representation. The next step is to promote the fields to the status of operators and impose the following non-vanishing, equal time commutation relations:
\begin{equation}
  \left[\phi(x), \Pi_{\phi}(y)\right] = \left[\phi^{\ast}(x), \Pi_{\phi^{\ast}}(y)\right] = i\delta(\x-{\bf y}).
\end{equation}
All other commutators of the field and conjugate momentum operators vanish. The operators $a(\p)$ and $b(\p)$ are the free particle and anti-particle annihilation operators respectively, whereas the operators $a^{\dagger}(\p)$ and $b^{\dagger}(\p)$ are the free particle and anti-particle creation operators. The vacuum state $|0 \ket$ is defined by $a(\p) | 0 \ket = b(\p) | 0 \ket = 0$. Note that the canonical quantization is performed in the interaction picture.

Expressing the creation and annihilation operators in terms of the fields and its derivatives, the commutation rules become
\begin{equation}
   \left[a({\bf p}), a^{\dagger}({\bf q})\right] = \left[b({\bf p}), b^{\dagger}({\bf q})\right] = \delta({\bf p}-{\bf q}).
\end{equation}

In the present work, we shall not be concerned with vacuum energy questions so we shall normal-order the creation and annihilation operators in the Hamiltonian. To obtain the Hamiltonian operator we integrate out the spatial coordinates from the Hamiltonian density (\ref{MOD_HAMILTONIAN}):
\begin{equation}
    \hat{H}(t) = \int d\x \, :{\cal H}(t, \x):,
\end{equation}
and express it in terms of the particle and anti-particle operators $a$, $b$, $a^\dagger$ and $b^\dagger$. This is straightforward for the free-field part of the Hamiltonian as can be seen in standard books on QFT. However, obtaining the analogous expressions for the interaction terms is tedious and this has been accomplished with the use of a computer code. 

The momentum operator is similarly obtained by normal-ordering and integrating out the spatial coordinates from the momentum density (\ref{MOMENTUM}):
\begin{equation}
  {\bf \hP} = \int d\x \, : \vec{\cal P}(\x):.
  \label{EQ:MO}
\end{equation}
Notice that since the momentum density operator does not involve the interaction terms, it is time independent. 

For the description of stationary few-particle bound states, it is convenient to switch to the Schr\"odinger picture where the time-dependence of the Hamiltonian is removed. The two pictures are related by 
\begin{equation}
  |\Psi_{I}(t) \ket =\e^{i \, H_{0}\, t} \, |\Psi_{S}(t) \ket
  \label{EQ:PICTURES}
\end{equation}
where $H_0$ is the free-field part of the Hamiltonian, \ie with interactions turned off. 

%% file: two.tex
\section*{Particle-Antiparticle State}
The simplest particle-antiparticle trial state can be expressed in terms of Fock-states as follows:
\begin{equation}
  |\Psi_2\ket = \int d\p_{1,2} \; F(\p_{1,2})\; a^{\dag} (\bp_1) \, b^{\dag} (\bp_2) \; | 0 \ket,
  \label{TRIAL_TWO}
\end{equation}
where $F(\bp_{1,2})$, is an adjustable coefficient function which is determined variationally. The subscript notation means that $d\p_{1,2} = d\p_{1} \, d\p_{2}$ and $F(\p_{1,2}) = F(\p_1, \p_2)$; it is used throughout this paper. 

\begin{figure}[t]
  \center{
    \includegraphics[scale=0.7]{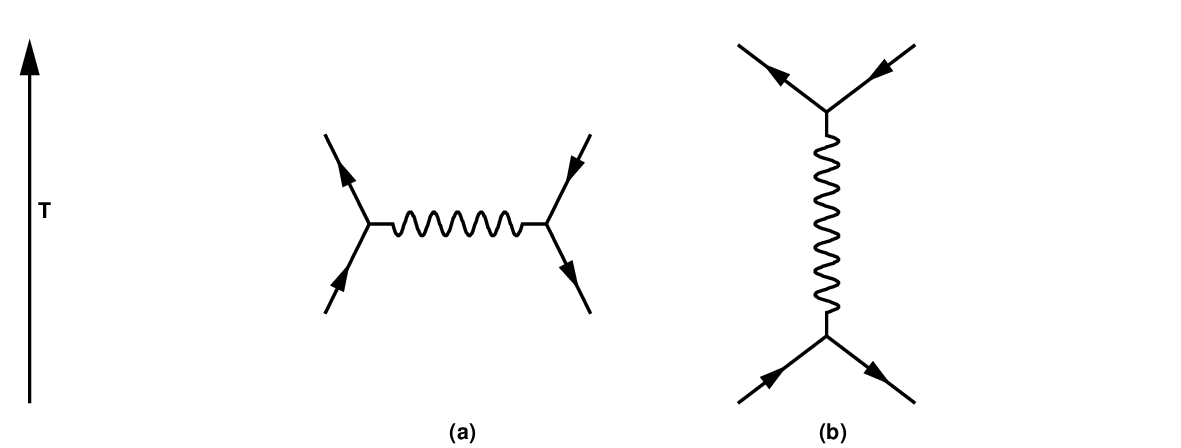}
  }
  \caption{The one-chion exchange (a) and virtual annihilation (b) diagrams of the Yukawa kernel equation \re{EQ:Y_22}.}
  \label{FIG:FD1}
\end{figure}

To implement the variational principle (\ref{EQ:VAR_PRIN}), the matrix element $\bra \Psi_2 | \, \hat{H} \, - E \, | \Psi_2 \ket$ is worked out and is varied with respect to the adjustable coefficient function $F^{\ast}$.  This leads to the following relativistic wave equation in momentum space describing a particle-antiparticle system:
\begin{equation}
  F(\p_{1,2}) \, \big(\omega_{\p_1} + \omega_{\p_2} - E\big) = \int d\p^{\prime}_{1,2} \; \cY_{2,2}(\p^{\prime}_{1,2}, \p_{1,2}) \, F(\p^{\prime}_{1,2}),
  \label{EQ2}
\end{equation}
where $\cY_{2,2}$ is the relativistic Yukawa particle-antiparticle interaction kernel (inter-particle ``momentum-space'' potential). In the interaction picture, it is given by
\begin{align}
  \cY_{2,2} (\bft{p}^{\prime}_{1,2}, \bft{p}_{1,2}) = & \; \frac{g^{2}}{8\,(2\pi)^{3}} \, \frac{\e^{\i \, (\omega_{\p^{\prime}_1} + \omega_{\p^\prime_2}-\omega_{\p_1}-\omega_{\p_2}) \, t}}{\ds\sqrt{\ds\omega_{\p^{\prime}_{1}}\ds\omega_{\p^{\prime}_{2}}\ds\omega_{\p_{1}}\omega_{\p_{2}}}} \, \ds\delta(\p^{\prime}_{1}+\p^{\prime}_{2}-\p_{1}-\p_{2}) \nonumber \\
  & \times \Bigg\{\frac{1}{\mu^{2}-(p^{\prime}_1-p_1)^{2}}+\frac{1}{\mu^{2}-(p^{\prime}_2-p_2)^{2}} + \frac{1}{\mu^{2}-(p_{1}+p_{2})^{2}} + \frac{1}{\mu^{2}-\ds(p^{\prime}_{1}+p^{\prime}_{2})^{2}}\Bigg\}.
  \label{EQ:Y_22}
\end{align}
Notice that the first two terms in the square brackets on the right hand side of equation (\ref{EQ:Y_22}) correspond to one-quantum exchange and the last two to virtual annihilation Feynman diagrams. These diagrams are shown in Figure \ref{FIG:FD1}.

The time dependence of the particle-antiparticle interaction kernel $\cY_{2,2}$ can be ``rotated away'' by the use of equation (\ref{EQ:PICTURES}). If we start with the trial state $| \Psi_2^\prime \ket = \e^{\i \, H_0 \, t} | \Psi_2 \ket$ then the corresponding interaction kernel in the Schr\"odinger picture will be time independent, \ie we obtain equation (\ref{EQ:Y_22}) with $t=0$. Therefore, the Schr\"odinger picture (time independent) interaction kernels can be found, in effect, by setting $t = 0$ in all interaction picture kernels. This is what we shall do in the remainder of the paper.
Unfortunately, the quantized version of the non-linear terms (\ref{MOD_HAM_C}) and (\ref{MOD_HAM_Q}) of the Hamiltonian density are not probed by the simple particle-antiparticle trial state (\ref{TRIAL_TWO}), \ie the matrix elements $\bra \Psi_2 | \, \hat{H}_{I_i} \, | \Psi_2 \ket$ vanish for $i = 2, 3$. To examine the effects of these terms we must use more elaborate  trial states than (\ref{TRIAL_TWO}). As it is, with the trial state (\ref{TRIAL_TWO}), the problem reduces to the scalar Yukawa model which has been discussed by Ding and Darewych~\cite{Ding:1999ed}. Nevertheless, it is useful to recount some details.

For the particle-antiparticle trial state (\ref{TRIAL_TWO}) to be an eigenstate of the momentum operator (\ref{EQ:MO}) we require that ${\bf \hP} \, |\Psi_2 \ket =  {\bf Q} \, |\Psi_2 \ket$. In the rest frame, ${\bf Q} = 0$, this requires that $F(\p_{1,2}) = \delta(\p_1 + \p_2) \, f(\p_1)$. The centre of mass motion separates and the relativistic momentum-space particle-antiparticle wave equation simplifies to
\begin{multline}
  f(\p) \, \big(2 \, \omega_\p - E\big) = \frac{g^{2}}{4\,(2\pi)^{3}} \int \frac{d\p^{\prime}}{\ds\omega_{\p^{\prime}}\ds\omega_\p} \, f(\p^{\prime}) \\ \times \, \Bigg\{\frac{1}{\mu^2+(\p^{\prime} - \p)^2 - (\omega_{\p^{\prime}} - \omega_{\p})^2} + \frac{1}{2}\left(\frac{1}{\mu^2 - 4 \, \omega^2_{\p}}\right) + \frac{1}{2}\left(\frac{1}{\mu^2 - 4 \, \omega^2_{\p^\prime}}\right)\Bigg\}.
  \label{EQ2A}
\end{multline}
The first term on the right hand side of equation (\ref{EQ2A}) corresponds to the one-``chion'' exchange (\ie one mediating field quantum exchange) while the second corresponds to the virtual annihilation Feynman diagrams. 

In the non-relativistic limit (\ie $\p^2 << m^2$) the Fourier transform of equation (\ref{EQ2A}) yields the expected Schr\"odinger equation in coordinate space for the relative motion of the particle-antiparticle system:
\begin{equation}
  -\frac{1}{m}\nabla^2 \, \psi(\x) + V(x) \, \psi(\x) = \varepsilon \, \psi(\x),
  \label{EQ:SCHROD}
\end{equation}
where $\varepsilon = E - 2 \, m$ and the potential energy depends on the particle-antiparticle distance: 
\begin{equation}
  V (x) = -\alpha_g \; \frac{\e^{-\mu x}}{x} + \frac{4 \, \pi \, \alpha_g}{\mu^2 - 4m^2}\delta(\x), 
  \label{PE2}
\end{equation}
where $\alpha_g = \ds\frac{g^2}{16 \, \pi m^2}$ is a dimensionless coupling constant. The first term, due to the one-chion exchange, is the usual Yukawa potential and is always attractive (\ie gravity-like) in this scalar theory. The second term, due to the virtual annihilation, can be either attractive or repulsive depending on the values of the parameters $\mu$ and $m$. It is a correction to the Yukawa potential and is a feature of the quantum field theory.

The relativistic equation (\ref{EQ2A}) is not analytically solvable, so approximation methods must be used. Perturbative and variational approximations are presented in the paper by Ding and Darewych~\cite{Ding:1999ed}, along with a comparison to results obtained using the ladder Bethe-Salpeter equation and various quasi-potential approximations. 

%% file: three.tex
\section*{Three Particle State}
To observe the effects of the non-linear terms of the Hamiltonian density (\ref{MOD_HAMILTONIAN}) on the inter-particle potential, we must consider trial states with either more particle content or more Fock-space components. We carry out the easier task first and consider a three identical particle trial state given by 
\begin{equation}
  |\Psi_3\ket = \int d\p_{1,2,3} \; F(\p_{1,2,3}) \; a^{\dag} (\bp_1) \, a^{\dag} (\bp_2) \, a^{\dag} (\bp_3) \; | 0 \ket,
  \label{TRIAL_THREE}
\end{equation}
where $F(\p_{1,2,3})$ is a three-particle function to be determined variationally. Note that this trial state can be taken to be an eigenstate of the momentum operator (\ref{EQ:MO}), $\ie {\bf \hP} \, |\Psi_3 \ket = {\bf Q} \,  |\Psi_2 \ket$ with the choice $F(\p_{1,2,3}) = \de(\p_1 + \p_2 + \p_3 - {\bf Q}) \, f(\p_{1,2})$, where ${\bf Q}$ is the constant total momentum of the three particle system. Consequently, the wavefunction will be of the form where the centre of mass motion is completely separable, just as we saw for the particle-antiparticle case. 
The addition of an extra third particle operator to the trial state significantly complicates the evaluation of the matrix element. However, the implicit symmetry of equation (\ref{TRIAL_THREE}) with respect to interchanges of momentum variables due to the identity of the particles makes the task easier. It permits us to carry out the evaluation of the matrix element for the three particle trial state in terms of the symmetrized function $F_S$:
\begin{equation}
  F_S(\p_{1,2,3}) = \sum_{i_1, i_2, i_3}^6 F(\p_{i_1,i_2,i_3}),
\end{equation}
where the summation is on the six permutations of the indices $1, 2$ and $3$. Making use of this symmetrization enables us to keep the arguments of the kernels non-permuted while storing all information about the symmetry under interchanges of the momentum variables in $F_S$. 

\begin{figure}[t]
  \center{
    \includegraphics[scale=0.6]{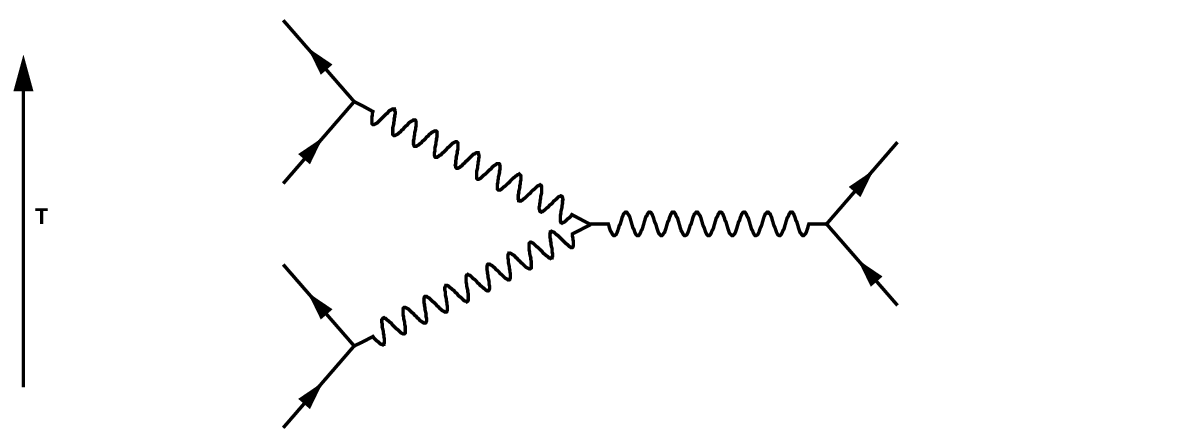}
  }
  \caption{The three-chion propagator vertex corresponding to the cubic interaction kernel $\cC_{3,3}$ equation (\ref{EQ:C_33}). The two propagators on the left should actually overlap (impossible to draw) such that they are perpendicular to the direction of time.}
  \label{FIG:FD2}
\end{figure}

We then calculate the matrix element $\bra \Psi_3 | \, \hat{H} \, - E \,| \Psi_3 \ket$, carry out the variational derivative with respect to $F^{\ast}$ and set it to zero (The complete expressions for the matrix element as well as other intermediate steps in the calculations are given in Appendix A). This yields the following relativistic momentum space integral wave equation for stationary states of three identical particles:
\begin{multline}
  F_S(\p_{1,2,3}) \, \big(\omega_{\p_1} + \omega_{\p_2} + \omega_{\p_3} - E\big) = \\ \int d\p^{\prime}_{1,2} \; \cY_{3,3}(\p^{\prime}_{1,2,3}, \p_{1,2,3}) \, F_S(\p^{\prime}_{1,2,3}) + \int d\p^{\prime}_{1,2,3} \; \cC_{3,3}(\p^{\prime}_{1,2,3}, \p_{1,2,3}) \, F_S(\p^{\prime}_{1,2,3}).
  \label{EQ3}
\end{multline}
The relativistic kernels of the interactions (\ie the relativistic momentum-space inter-particle potentials) are given by 
\begin{align}
  \cY_{3,3}(\p^{\prime}_{1,2,3}, \p_{1,2,3}) =  & -\frac{g^2}{8\,(2\pi)^{3}} \sum_{i_1, i_2, i_3}^6 \frac{\delta(\p^{\prime}_1 + \p^{\prime}_2 - \p_{i_1} - \p_{i_2}) \, \delta(\p^{\prime}_3 - \p_{i_3})}{\ds\sqrt{\omega_{\p^{\prime}_{1}}\omega_{\p^{\prime}_{2}}\omega_{\p_{i_1}}\omega_{\p_{i_2}}}} \, \Bigg\{\frac{1}{\mu^2-(p^{\prime}_1 - p_{i_1})^2}\Bigg\}, 
  \label{EQ:Y_33} \\
  \cC_{3,3}(\p^{\prime}_{1,2,3}, \p_{1,2,3}) = & - \frac{g^{3}\eta}{8(2\pi)^{6}} \sum^{6}_{i_1, i_2, i_3} \frac{\delta(\p^{\prime}_1 + \p^{\prime}_2 +\p^{\prime}_3 - \p_{i_1} - \p_{i_2} - \p_{i_3})}{\ds\sqrt{\omega_{\p^{\prime}_{1}}\omega_{\p^{\prime}_{2}}\omega_{\p^{\prime}_{3}}\omega_{\p_{i_1}}\omega_{\p_{i_2}}\omega_{\p_{i_3}}}} \nonumber \\
  & \F! \times \Bigg\{\frac{1}{\mu^2-(p^{\prime}_1 + p^{\prime}_2 - p_{i_1} - p_{i_2})^2}\frac{1}{\mu^2-(p^{\prime}_1 - p_{i_1})^2}\frac{1}{\mu^2-(p^{\prime}_2 - p_{i_2})^2}\Bigg\}, 
  \label{EQ:C_33}
\end{align}
where the summation is on the six permutations of the three indices $1, 2$ and $3$. As is evident from the covariant factors on the RHS of equation (\ref{EQ:Y_33}), the kernel $\cY_{3,3}$ corresponds to the three inter-particle one-chion exchange interactions. There are no virtual annihilation terms since only particle operators (and no antiparticle operators) are present in the trial state (\ref{TRIAL_THREE}). The kernel $\cC_{3,3}$, equation (\ref{EQ:C_33}), corresponds to the non-linear interaction term $\cH_{I_2}$ (details are given in Appendix A). It is similarly evident, from the covariant factors on the RHS of equation (\ref{EQ:C_33}), that the Feynman diagram corresponding to the cubic kernel $\cC_{3,3}$ contains a three-chion propagator vertex and is shown in Figure \ref{FIG:FD2}. Note that the three particle trial state does not probe the quartic interaction term $\cH_{I_3}$ equation (\ref{MOD_HAM_Q}), \ie $\bra \Psi_3 | \, \hat{H}_{I_3} \, | \Psi_3 \ket = 0$.
Equation (\ref{EQ3}) (with its kernels equations (\ref{EQ:Y_33}) and (\ref{EQ:C_33})) is one of our principal results. It is a relativistic wave equation for stationary states of a system of three identical scalar particles. It can describe purely bound states of the three particles or elastic scattering among them, but not processes that involve the emission or absorption of chions. The relativistic kinematics (\ie kinetic energy) of the system is described without approximation, but the relativistic dynamics (\ie potential energy) is described only at the level of one-chion exchange between the particle pairs by the relativistic kernel $\cY_{3,3}$, along with a three-particle first iterative order correction $\cC_{3,3}$ due to the non-linear interaction term $\cH_{I_2}$. 

The solution of the relativistic three particle equation (\ref{EQ3}) (even the non-relativistic limit, which we discuss below) is a challenging task which we shall not undertake in this work. No exact solutions are possible, therefore approximation methods (perturbative, variational, numerical or other) must be used. This shall be left for a future investigation. 

It is of interest and instructive to consider the non-relativistic limit of equation (\ref{EQ3}), in particular the coordinate representation of the interactions. Moreover, the non-relativistic description can be used for systems of heavy particles. 

The non-relativistic limit of equations (\ref{EQ3})-(\ref{EQ:C_33}), just as for the particle-antiparticle system, is obtained by assuming $\p^2 << m^2$ and Fourier-transforming to coordinate space. From equation (\ref{EQ:Y_33}), the non-relativistic inter-particle potential for the three particle system due to the Yukawa interaction is, as expected,  
\begin{equation}
  V_Y(\x_{1,2,3}, \mu) = V_Y(x_{ij}, \mu) = - \, \alpha_g \Bigg\{\frac{\e^{- \mu x_{12}}}{x_{12}} + \frac{\e^{- \mu x_{13}}}{x_{13}} + \frac{\e^{- \mu x_{23}}}{x_{23}}\Bigg\}.
  \label{EQ:V_Y123} 
\end{equation}
Unlike in the particle-antiparticle case (\ref{PE2}), no virtual annihilation delta function contributions appear in this expression since we are dealing with a system of three identical particles only (\ie there are no antiparticles). 

The non-relativistic cubic potential term that follows from equation (\ref{EQ:C_33}) is
\begin{equation}
  V_C(\x_{1,2,3}, \mu) = - \, \pi^3\alpha_{\eta} \int d\q_{1,2,3} \, \frac{\prod_{i}^{3}\e^{-\i \q_{i}\cdot\x_{i}} \, \delta(\q_1+\q_2+\q_3)}{\left(\mu^2+\q_1^2\right)\left(\mu^2+\q_2^2\right)\left(\mu^2+\q_3^2\right)},
  \label{EQ:V_C123}
\end{equation}
where $\alpha_{\eta} = \ds\frac{3 \, g^3\eta}{256 \, \pi^6m^3}$ is a coupling constant with dimensions of mass (see Appendix A for details). $V_C$ can be simplified to a three-dimensional quadrature (see Appendix A for details):
\begin{equation}
  V_C(\x_{1,2,3}, \mu) = - \, \pi^3 \alpha_\eta \, \int d\x \, \frac{\e^{-\mu|\x_1+\x|}}{|\x_1+\x|} \, \frac{\e^{-\mu|\x_2+\x|}}{|\x_2+\x|} \, \frac{\e^{-\mu|\x_3+\x|}}{|\x_3+\x|}.
  \label{EQ:V_C123A}
\end{equation}
This is an overall well behaved convergent integral for any $\mu > 0$. However, it cannot be evaluated analytically in general (at least we do not know how to do so). 
The expression for $V_C$, equation (\ref{EQ:V_C123A}), with the substitution $\v = \x + \x_1$, can be written as 
\begin{equation}
  V_C(\x_{1,2,3}, \mu) = - \, \pi^3 \alpha_\eta \int d\v \, \frac{\e^{-\mu |\v|}}{|\v|} \, \frac{\e^{-\mu |\v + \x_{21}|}}{|\v + \x_{21}|} \, \frac{\e^{-\mu |\v + \x_{31}|}}{|\v + \x_{31}|},
  \label{EQ:V_C123B}
\end{equation}
where $\x_{21} = \x_2 - \x_1$ and $\x_{31} = \x_3 - \x_1$. As shown in Appendix C, we can also express $V_C$ in the form
\begin{equation}
  V_C(\x_{ij}, \mu) = - \, \pi^3 \alpha_{\eta} \int_{0}^{\infty}d\beta_{1,2,3} \; \frac{\e^{-\mu^{2}(\beta_{1}+\beta_{2}+\beta_{3})}}{(\beta_{1}\beta_{2}+\beta_{1}\beta_{3}+\beta_{2}\beta_{3})^{3/2}} \; \exp\left(-\frac{\beta_{1}\x_{21}^{2}+\beta_{2}\x_{31}^{2}+\beta_{3}\x_{32}^{2}}{4\left(\beta_{1}\beta_{2}+\beta_{1}\beta_{3}+\beta_{2}\beta_{3}\right)}\right),  
  \label{EQ:V_C123C}
\end{equation}
which shows explicitly that $V_C$ depends only on inter-particle distances $x_{ij} = x_{ji} =|\x_i - \x_j|$. In light of equation (\ref{EQ:V_C123C}), we conclude that the total potential energy function $V = V_Y + V_C$ depends only on the inter-particle distances, and is invariant under 3D rotations and translations of coordinates as we would expect of a closed system.  Therefore, we shall also use the notation $V_C(x_{ij}, \mu)$ in what follows. In addition, equation (\ref{EQ:V_C123C}) is suitable for numerical evaluations of $V_C$ for arbitrary values of $x_{ij}$. 

Unfortunately, it is not possible to plot inter-particle potentials of three independent variables $x_{12}$, $x_{23}$ and $x_{13}$, even though they can be worked out numerically for arbitrary $x_{ij}$. For this reason, we will plot one and two dimensional sections to get some idea of the shape of the cubic term $V_C$.

We proceed to examine two particular cases for which $V_C$ can be evaluated analytically. Then, we show surface plots of $V_C$ for arbitrary $x_{ij}$ which we calculated numerically. 
\subsubsection*{Case 1: $V(\x_{1,2,2}, \mu) = V(x_{21}, \mu)$}
\begin{figure}[ht!] 
  \begin{center}
    \includegraphics[scale = 0.60]{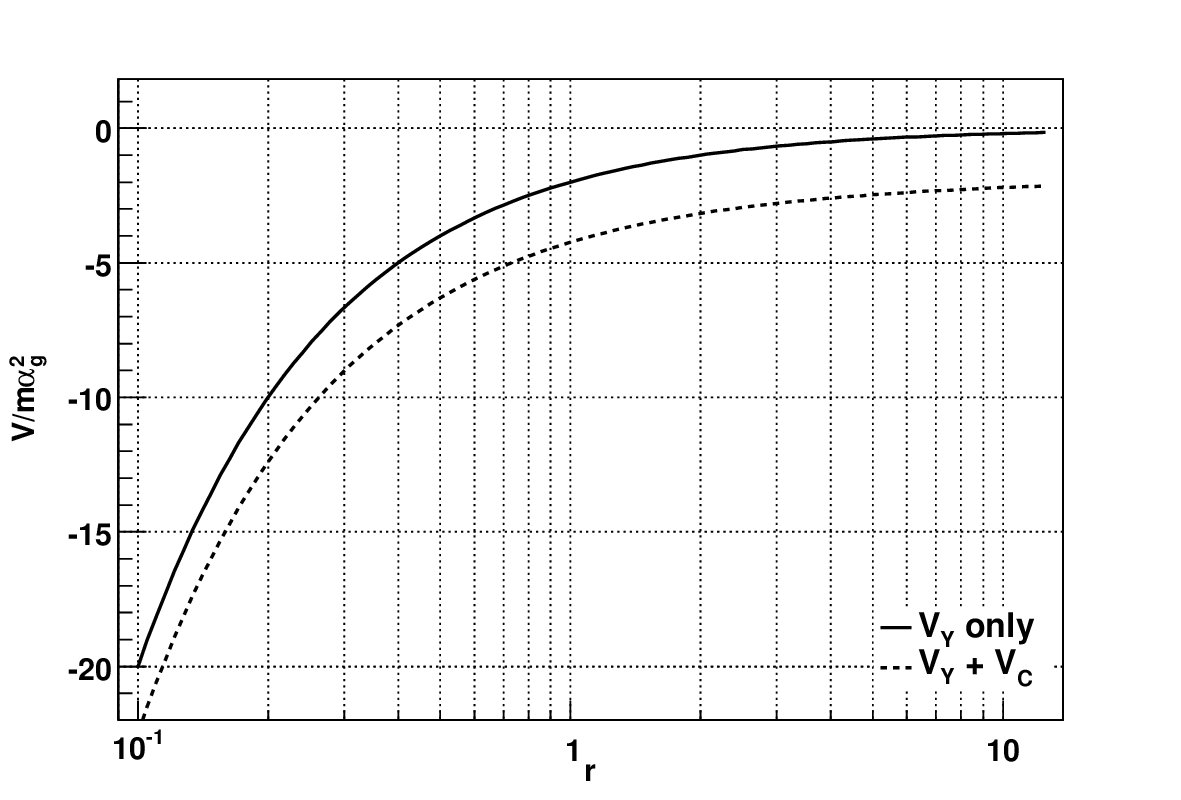}
    \caption{The inter-particle potential $V(r, M) \, / \, m\alpha^2_g$ of equation (\ref{EQ:V_122MU}) as a function of $r = x_{21}\, m\alpha_g$ for $M = 10^{-10}$ with $\kappa_1 = 0$ (i.e. $V_Y$ only) and $\kappa_1 = 0.1$ (i.e. $V_Y + V_C$).}
    \label{FIG:V_122SEPARATE}
    \includegraphics[scale = 0.60]{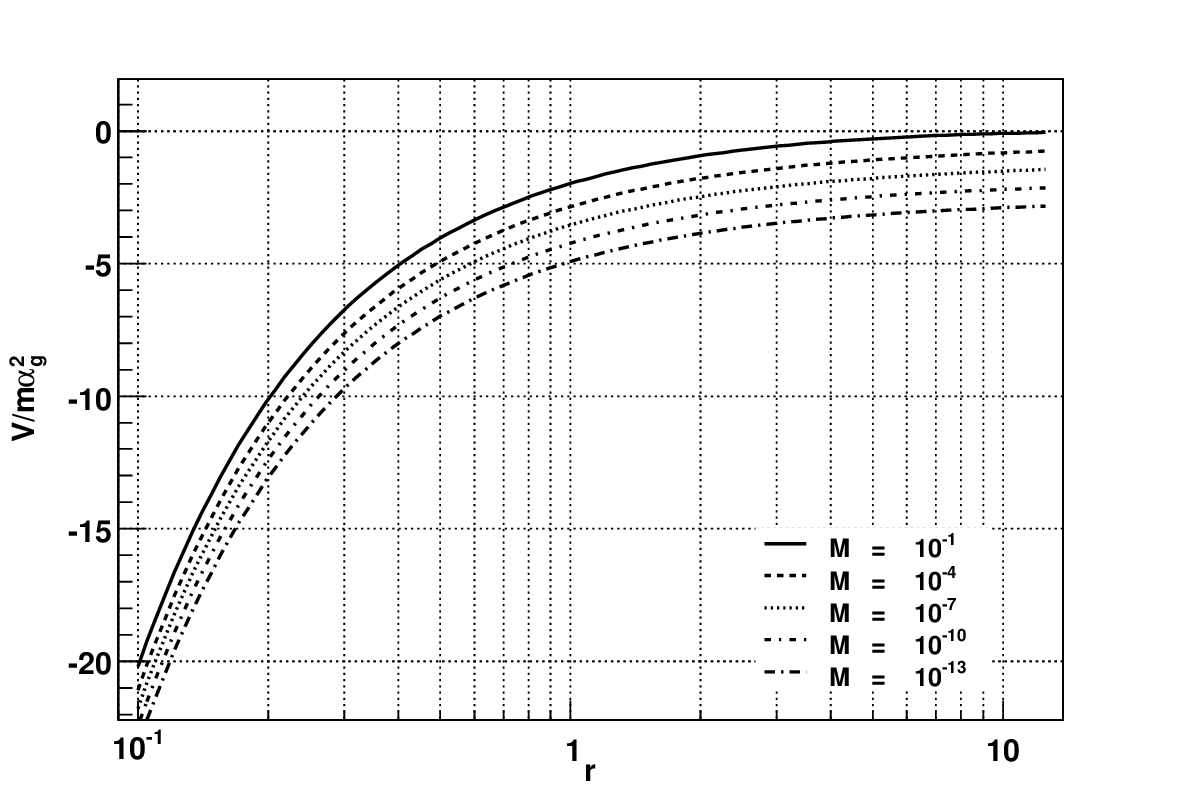}
    \caption{The inter-particle potential $V(r, M) \, / \, m\alpha^2_g$ of equation (\ref{EQ:V_122MU}) as a function of $r = x_{21}\, m\alpha_g$ for $\kappa_1 = 0.1$ and various values of $M$.}
    \label{FIG:V_122MU}
  \end{center}
\end{figure}

The first case we examine is when $\mu > 0$  and $\x_2 = \x_3$ so that there is only one inter-particle distance, $x_{21} = x_{12} = |\x_2 - \x_1| = |\x_3 - \x_1|$, to be concerned with. The integral for $V_C$, equation (\ref{EQ:V_C123B}), in this case is expressible in terms of the special function, the exponential integral, defined by 
\begin{equation}
  \E1(z) = \int^{\infty}_z \frac{\e^{-t}}{t} \, dt,
  \label{EQ:EXPINT}
\end{equation}
namely
\begin{equation}
  V_C(\x_{1,2,2} ,\mu) = V_C(x_{21}, \mu) = -\frac{2 \, \pi^4\alpha_{\eta}}{x_{21} \mu} \Bigg\{\e^{-x_{21} \mu}\ds\Big(\ln\left(3\right) - \E1\left(x_{21}\mu\right)\Big) + \e^{x_{21}\mu} \, \E1\left(3 \, x_{21}\mu\right) \Bigg\}.
  \label{EQ:V_C122MU}
\end{equation}
We choose to express length in units of the Bohr radius $\ds\frac{1}{m\alpha_g}$ and, correspondingly, the energy in units of $m\alpha^2_g$. Equation (\ref{EQ:V_C122MU}) is then written in terms of the dimensionless variables $r = x_{21} \, m \, \alpha_g$ and $M=\ds\frac{\mu}{m\alpha_g}$. Suppressing the singularity at $\x_2 = \x_3$ in the Yukawa term (\ref{EQ:V_Y123}), the total inter-particle potential $V =V_Y + V_C$ for the three particle trial state in the case when $\mu > 0$ and $\x_2 = \x_3$ is 
\begin{multline}
  V(\x_{1,2,2}, \mu > 0) = V(r,M) \\
  = m \, \alpha^{2}_g \; \Bigg\{ - \, 2 \, \frac{\e^{- r \, M}}{r} - \frac{\kappa_1}{2}\ds\left(\frac{\e^{-r \, M}}{r \, M} \, \ln(3) 
  - \frac{\e^{-r \, M}}{r \, M} \, \E1(rM)
  + \frac{\e^{r \, M}}{r \, M} \, \E1(3 \, r \, M) \right) \Bigg\},
  \label{EQ:V_122MU}
\end{multline}
where $\kappa_1 = \ds\frac{4\pi^4\alpha_\eta}{m\alpha^2_g}=\ds\frac{12\eta}{g}$ is a dimensionless constant. Figures (\ref{FIG:V_122SEPARATE}) and (\ref{FIG:V_122MU}) are plots of $V(r,M)/ m \, \alpha_g^2$, equation (\ref{EQ:V_122MU}), as a function of $r$ for the different values of $M$, with $\kappa_1 = 0$ and $\kappa_1 = 0.1$ respectively. It is apparent that the contribution of $V_C$ lowers the overall potential with decreasing values of $M$.

The behaviour of the inter-particle potential can be better understood by looking at the two extreme cases of the separation distance $r$. For small $r$, expanding equation (\ref{EQ:V_122MU}), we obtain
\begin{equation}
  V(r,M) \approx - \frac{2}{r} + \kappa_1 \, \ln( 3\, r \, M) + C_M, 
\end{equation}
where $C_M = 2 \, M + \kappa_1 \, (\gamma - 1)$ is a finite constant with $\gamma = 0.5772$ being the Euler's constant. The leading $-2/r$ term in the above expansion indicates that the inter-particle potential is dominated by the Yukawa term $V_Y$ for small $r$. To see the large $r$ behaviour, we determine analytically the ratio $V_C / V_Y$: 
\begin{equation}
  \frac{V_C}{V_Y} = \frac{\kappa_1}{4M} \, \Big( \ln(3) - \E1(r \, M) + \e^{2 \, r \, M} \, \E1( 3 \, r \, M) \Big).
\end{equation}
The large $r$ limit of this ratio is the constant $\ds\frac{\ln(3) \, \kappa_1}{4M}$ which implies that the cubic term $V_C$ has the identical large $r$ dependence as the Yukawa term $V_Y$. From this we see that the effect of $V_C$ is to strengthen the attractive Yukawa-like potential, i.e. to increase the binding energies of bound states.

\subsubsection*{Case 2: $V(x_{21} = x_{31} = x_{23} = \Delta, \mu) = V(\Delta, \mu)$}

\begin{figure}[t] 
  \begin{center}
    \includegraphics[scale = 0.7]{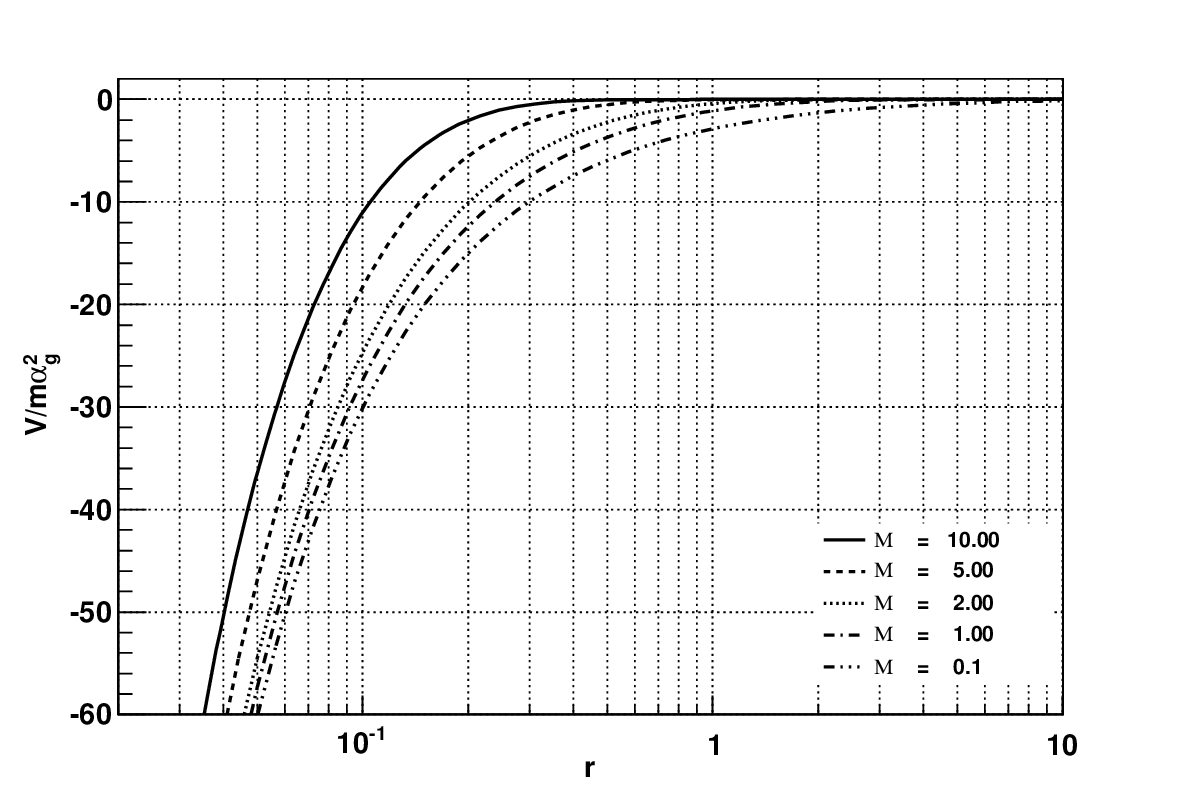}
    \caption{The inter-particle potential $V(r, M)\, / \, m\alpha^2_g$ of equation (\ref{EQ:V_DMU}) for $\kappa_1 = 0.1$ and the indicated values of  $M = \ds\frac{\mu}{m\alpha_g}$, where $r = \Delta \, m \alpha_g$ and $\Delta = x_{21} = x_{31} = x_{23}$.}
      \label{FIG:V_DMU}
  \end{center}
\end{figure}

Another case for which $V_C$ can be expressed in closed from is when the coordinates are at the vertices of an equilateral triangle (see Appendix A for details). With the previous choice of the units of length and energy, we have
\begin{multline}
   V(\Delta, \mu ) = V(r, M) 
   \\ = - \, m \, \alpha_g^2 \,  \Bigg\{ \, 3 \, \frac{\e^{-r M}}{r} + \frac{\kappa_1}{4\pi}\int_0^\infty d\beta_{1,2,3} \; \frac{\e^{-M^{2}(\beta_{1}+\beta_{2}+\beta_{3})}}{(\beta_{1}\beta_{2}+\beta_{1}\beta_{3}+\beta_{2}\beta_{3})^{3/2}} \, \exp\left(-\frac{r^2 \, (\beta_1 + \beta_2 + \beta_3)} {4(\beta_1\beta_2 + \beta_1\beta_3 + \beta_2\beta_3)}\right) \Bigg\}
  \label{EQ:V_DMU}
\end{multline}
where $r = \Delta \, m \alpha_g$ is the dimensionless inter-particle distance, $\Delta = |\x_2 - \x_1| = |\x_3 - \x_1| = |\x_3 - \x_2|$ and $M = \ds\frac{\mu}{m\alpha_g}$ is the dimensionless mass parameter of the mediating field. Here, we evaluated $V_C$ numerically. The numerical integration was performed with the GNU Scientific Library. A plot of equation (\ref{EQ:V_DMU}) is given in Figure \ref{FIG:V_DMU}. The curves show that $V(r,M)$ approaches zero asymptotically from below for all $M > 0$.

\subsubsection*{Case 3: Arbitrary inter-particle distances}
\begin{figure}
  \begin{center}$ 
    \begin{array}{cc}
      \includegraphics[scale = 0.35]{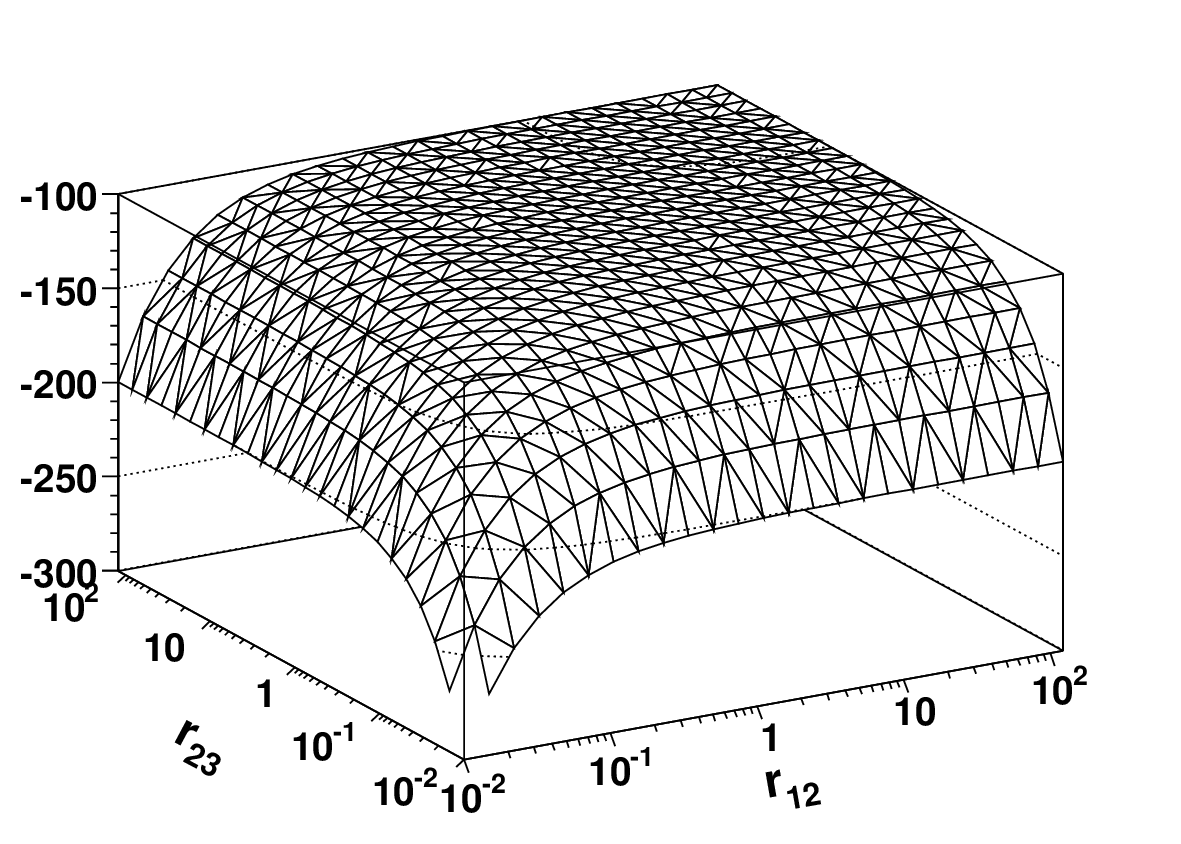} & \includegraphics[scale = 0.35]{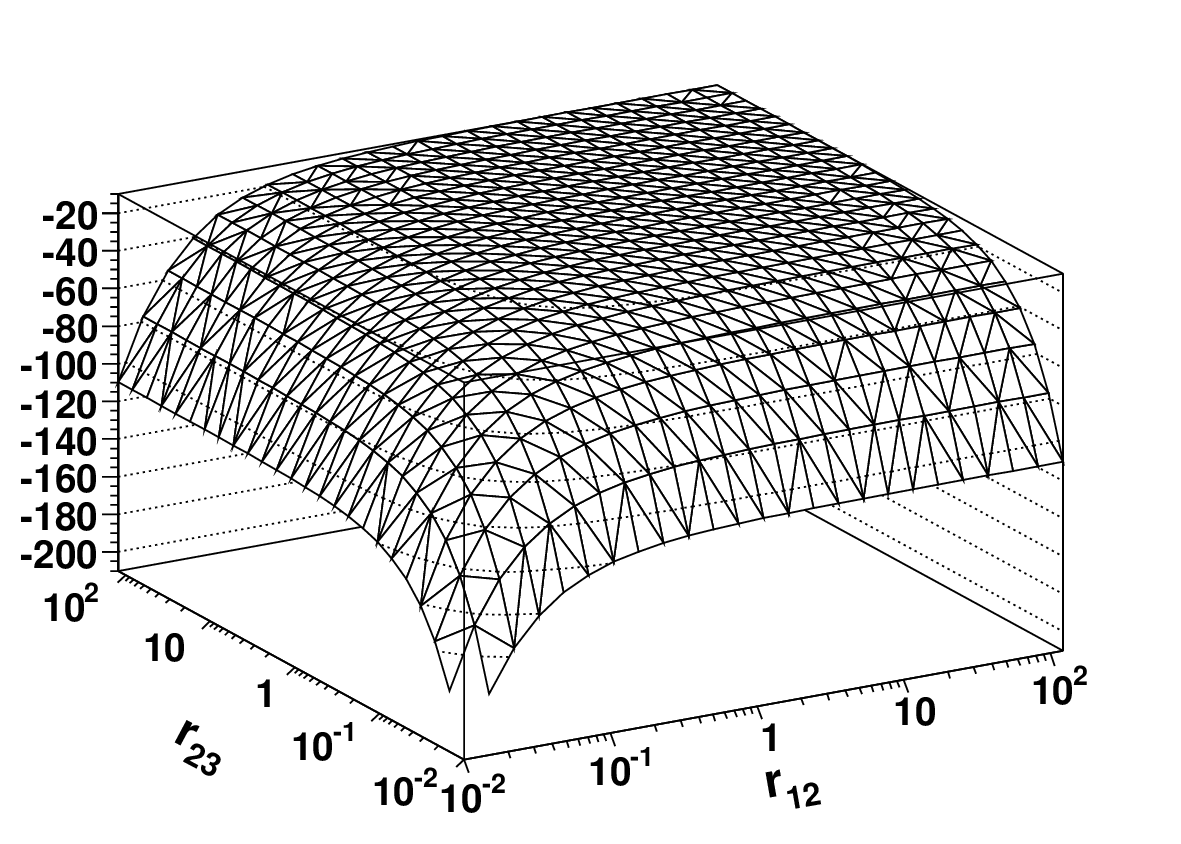} \\
      \includegraphics[scale = 0.35]{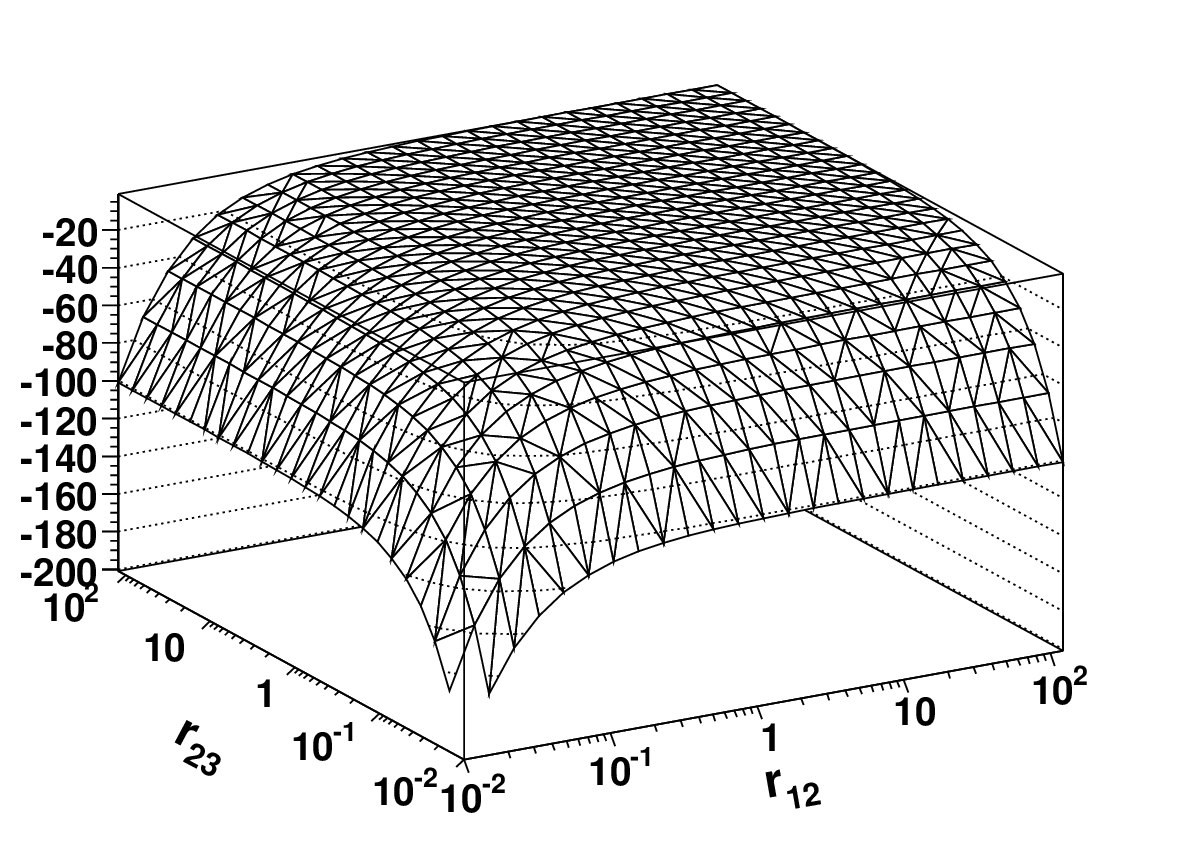} & \includegraphics[scale = 0.35]{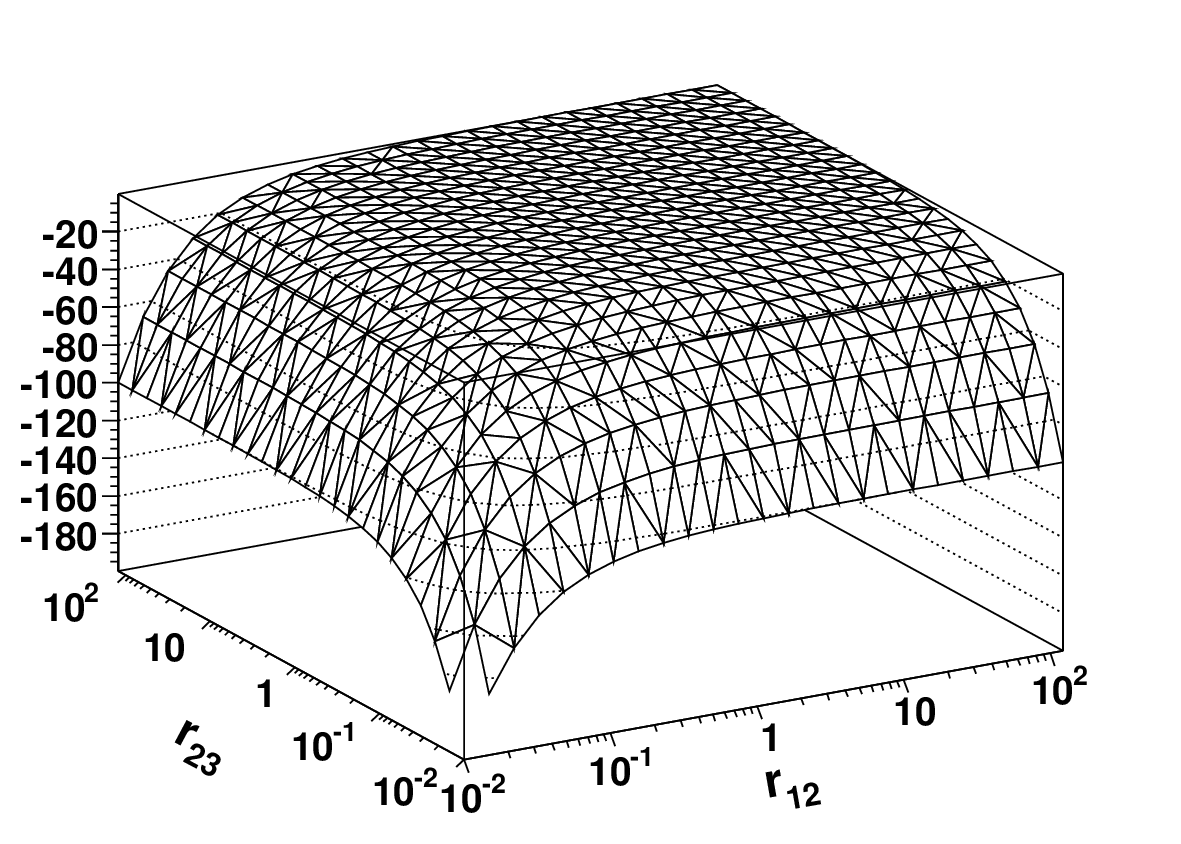} 
    \end{array}$
  \end{center}
  \caption{The inter-particle potential $V(r_{ij}, M)\, / \, m\alpha^2_g$ of equation (\ref{EQ:V_arbitrary}) for $\kappa_1 = 0.1$, $M = \ds\frac{\mu}{m\alpha_g} = 0.1$ and $r_{13} = 0.01 (\tx{top left}), 0.1(\tx{top right}), 1.0 (\tx{bottom left}), 10 (\tx{bottom right})$ where $r_{ij} = \Delta_{ij} \, m \alpha_g$.}
  \label{FIG:V_surface}
\end{figure}

Finally, we examine the inter-particle potential for arbitrary inter-particle distances. There are no analytical solutions available for $V_C$, equation (\ref{EQ:V_C123A}), and the potential is calculated numerically with the help of the GNU Scientific Library. Using the parametrization given by equation (\ref{EQ:V_C123C}), we obtain
\begin{multline}
  V(x_{ij}, \mu )  = V(r_{ij}, M) = - \, m \, \alpha_g^2 \, \Bigg\{ \, \frac{\e^{-r_{12} M}}{r_{12}} + \frac{\e^{-r_{23} M}}{r_{23}} + \frac{\e^{-r_{13} M}}{r_{13}}  \\
  + \frac{\kappa_1}{4 \, \pi}\int_0^\infty  \, d\beta_{1,2,3} \; \frac{\e^{-M^{2}(\beta_{1}+\beta_{2}+\beta_{3})}}{(\beta_{1}\beta_{2}+\beta_{1}\beta_{3}+\beta_{2}\beta_{3})^{3/2}} \, \exp\left(-\frac{r_{12}^2 \beta_1 + r_{23}^2 \beta_2 + r_{13}^2\beta_3} {4(\beta_1\beta_2 + \beta_1\beta_3 + \beta_2\beta_3)}\right) \Bigg\} 
  \label{EQ:V_arbitrary}
\end{multline}
where $r_{ij} = x_{ij} \, m \alpha_g$ are the dimensionless inter-particle distances and $x_{ij} = |\x_i - \x_j|$. Four surface plots of equation (\ref{EQ:V_arbitrary}) for different values of $r_{13}$ are given in Figure \ref{FIG:V_surface}. Note how the inter-particle potential rises with increasing separations among particles but does not cross the zero plane. 

We see that the cubic $\cH_{I_2}$ interaction term of the Hamiltonian affects the inter-particle interaction potential $V$ through the additional attractive term $V_C$. As a consequence, the energy spectrum for a three particle system will be lowered by the inclusion of the cubic term $\cH_{I_2}$.

%% file: four.tex
\section*{Four Particle State}
Since the three particle trial state (\ref{TRIAL_THREE}) does not probe the quartic $\cH_{I_3}$ interaction term of the Hamiltonian (\ie $\bra \Psi_3 | H_3 | \Psi_3 \ket = 0$), we consider a system of four identical particles and examine how both non-linear terms of the Hamiltonian, $\cH_{I_2}$ and $\cH_{I_3}$, affect the inter-particle potential. 

\begin{figure}[t]
  \center{
    \includegraphics[scale=0.6]{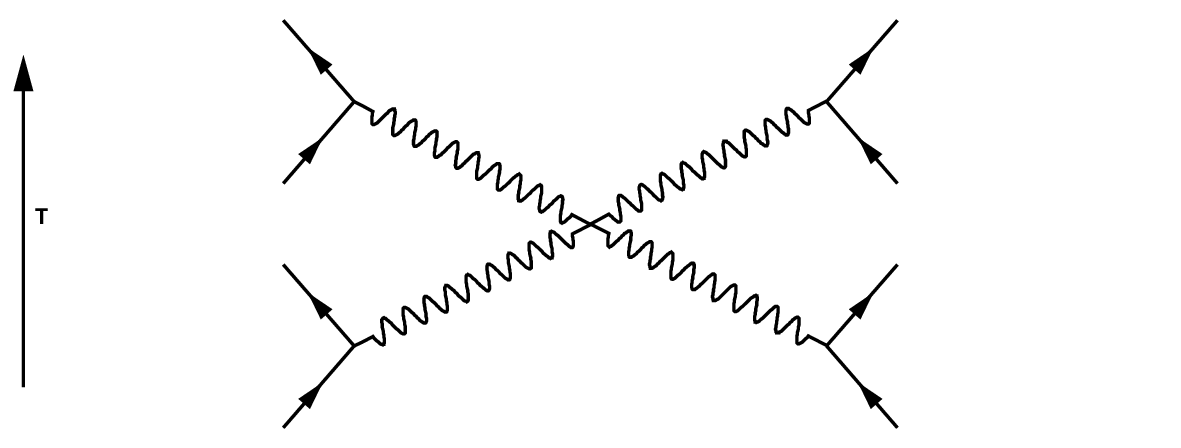}
  }
  \caption{The four-chion propagator vertex corresponding to the quartic interaction kernel $\cQ_{4,4}$ equation (\ref{EQ:Q44}). The two propagators on both sides should actually overlap (impossible to draw) such that they are perpendicular to the direction of time.}
  \label{FIG:FD3}
\end{figure}

The four identical particle trail state analogous to equation (\ref{TRIAL_THREE}) is given by 
\begin{equation}
    |\Psi_{4}\ket = \ds \int d\p_{1..4} \; F(\p_{1..4}) \; a^{\dagger}(\p_1) \, a^{\dagger}(\p_2) \, a^{\dagger}(\p_3) \, a^{\dagger}(\p_4) \, |0\ket.
    \label{TRIAL_FOUR}
\end{equation}
For this four particle trial state to be an eigenvector of the momentum operator (\ref{EQ:MO}), we require that $\hat{\bf P} \, | \Psi_4 \ket = {\bf Q} \,| \Psi_4 \ket$ which can be achieved with the choice $F(\p_{1..4}) = \de(\p_1 + \p_2 + \p_3 + \p_4 - {\bf Q}) \, f(\p_{1,2,3})$. Thereupon, similar to the particle-antiparticle and the three particle cases, the wavefunctions will be of the form where the centre of mass motion is completely separable for this identical particle system. We shall write everything in terms of the completely symmetrized function $F_S$ because the trial state (\ref{TRIAL_FOUR}) is completely symmetric under interchanges of the momentum variables. The symmetrized function is 
\begin{equation}
  F_S(\p_{1..4})  = \sum_{i_1, i_2, i_3, i_4}^{24} \,   F(\p_{i_1, i_2, i_3, i_4}), 
\end{equation}
where the summation is on the 24 permutations of the indices $1, 2, 3$ and $4$. We calculate the matrix element $\bra \Psi_4 | \, \hat{H} \, - E \,| \Psi_4 \ket$, work out the variational derivative with respect to $F^{\ast}$ and set it to zero (see Appendix B for details). This leads to the following four identical particle relativistic equation for the function $F_S$ (in momentum space): 
\begin{multline}
  F_S(\p_{1..4}) \, \big(\omega_{\p_1} + \omega_{\p_2} + \omega_{\p_3} + \omega_{\p_4} - E\big) = \int d\p^{\prime}_{1,2} \; \cY_{4,4}(\p^{\prime}_{1..4}, \p_{1..4}) \, F_S(\p^{\prime}_{1..4}) \\ + \int d\p^{\prime}_{1..4} \; \cC_{4,4}(\p^{\prime}_{1..4}, \p_{1..4}) \, F_S(\p^{\prime}_{1..4}) + \int d\p^{\prime}_{1..4} \; \cQ_{4,4}(\p^{\prime}_{1..4}, \p_{1..4}) \, F_S(\p^{\prime}_{1..4}).
  \label{EQ4}
\end{multline}
The Yukawa and cubic interaction kernels $\cY_{4,4}$ and $\cC_{4,4}$ are similar in structure to those of the three particle trial state except there is dependence on an extra momentum coordinate (see Appendix B). The relativistic quartic interaction kernel is 
\begin{align}
  \cQ_{4,4}&(\p^{\prime}_{1..4}, \p_{1..4}) = \frac{g^4\sigma}{16}\frac{1}{(2\pi)^9} \, \sum^{24}_{i_1, i_2, i_3, i_4} \, \frac{ \de(\p^{\prime}_1 + \p^{\prime}_2 + \p^{\prime}_3 + \p^{\prime}_4 - \p_1 -\p_2 - \p_3 -\p_4)}{\ds\sqrt{\omega_{\p^{\prime}_1}\omega_{\p^{\prime}_2}\omega_{\p^{\prime}_3}\omega_{\p^{\prime}_4}\omega_{\p_1}\omega_{\p_2}\omega_{\p_3}\omega_{\p_4}}} \nonumber \\
  & \times \Bigg\{\frac{1}{\mu^2-(p^{\prime}_1+p^{\prime}_2+p^{\prime}_3-p_{i_1}-p_{i_2}-p_{i_3})^2}\frac{1}{\mu^2-(p^{\prime}_{1}-p_{i_1})}\frac{1}{\mu^2-(p^{\prime}_2-p_{i_2})^2}\frac{1}{\mu^2-(p^{\prime}_3-p_{i_3})^2}\Bigg\},
  \label{EQ:Q44}
\end{align}
where the summation is on the 24 permutations of the indices $1, 2, 3$ and $4$. This contribution to the potential corresponds to a four-chion propagator vertex shown in Figure \ref{FIG:FD3}. 

It is not our intent to solve the four-particle relativistic equation (\ref{EQ4}) in this paper. No exact solutions are possible, just as with the three-particle relativistic equation (\ref{EQ3}). Instead, we proceed to examine the non-relativistic limit of the interactions in the coordinate representation. 

In the non-relativistic limit, the Yukawa and cubic kernel reduce just as their three particle trial state counter-parts. There are six Yukawa interaction terms which can be worked out in analytical form:
\begin{equation}
  V_Y(\x_{1..4}, \mu)= - \alpha_g \Bigg\{\frac{\e^{-\mu x_{12}}}{x_{12}}+\frac{\e^{-\mu x_{13}}}{x_{13}}+\frac{\e^{-\mu x_{14}}}{x_{14}}+\frac{\e^{-\mu x_{23}}}{x_{23}}+\frac{\e^{-\mu x_{24}}}{x_{24}}+\frac{\e^{-\mu x_{34}}}{x_{34}}\Bigg\},
\end{equation}
where $\alpha_g = \ds\frac{g^2}{16 \, \pi \, m^2}$ is the dimensionless coupling constant and $x_{ij} = x_{ji} = |\x_j - \x_i|$ are the inter-particle distances as before. 
The cubic interaction kernel for the non-relativistic four particle case reduces to a sum of four terms, \ie one for every three-way interaction:
\begin{multline}
  V_C(\x_{1..4}, \mu) = -\alpha_{\eta} \, \int d\bft{q}_{1..4}\, \Bigg\{\frac{\prod_{i}^{4}\e^{-\i \bft{q}_{i}\cdot\bft{x}_{i}} \, \delta(\bft{q}_1+\bft{q}_2+\bft{q}_3) \, \delta(\bft{q}_{4})}{\left(\mu^2+\bft{q}_{1}^2\right)\left(\mu^2+\bft{q}_{2}^2\right)\left(\mu^2+\bft{q}_{3}^2\right)} +  \frac{\prod_i^4 \e^{-\i \bft{q}_i\cdot\bft{x}_i} \, \delta(\bft{q}_1+\bft{q}_2+\bft{q}_4) \, \delta(\bft{q}_3)}{\left(\mu^2+\bft{q}_1^2\right)\left(\mu^2+\bft{q}_2^2\right)\left(\mu^2+\bft{q}_4^2\right)} \\
  + \frac{\prod_{i}^{4}\e^{-\i \bft{q}_{i}\cdot\bft{x}_{i}} \, \delta(\bft{q}_1+\bft{q}_3+\bft{q}_4) \, \delta(\bft{q}_{2})}{\left(\mu^2+\bft{q}_1^2\right)\left(\mu^2+\bft{q}_3^2\right)\left(\mu^2+\bft{q}_4^2\right)} +  \frac{\prod_i^4 \e^{-\i \bft{q}_i\cdot\bft{x}_i} \, \delta(\bft{q}_2+\bft{q}_3+\bft{q}_4) \, \delta(\bft{q}_1)}{\left(\mu^2+\bft{q}_2^2\right)\left(\mu^2+\bft{q}_3^2\right)\left(\mu^2+\bft{q}_4^2\right)} \Bigg\} 
  \label{EQ:V_C1234},
\end{multline}
where $\alpha_{\eta}=\ds\frac{3 \, g^3\eta}{256 \, \pi^6 m^3}$ is a coupling constant with dimensions of mass (see Appendix B for details). The delta function with a single momentum variable in every term indicates that the particle carrying that momentum is a ``spectator'' of the three-way interaction. To obtain this expression we followed similar steps as those leading to equation (\ref{EQ:V_C123}). 
The quartic interaction kernel for the four particle system in the non-relativistic limit reduces to
\begin{equation}
  V_Q(\x_{1..4}) = \alpha_{\sigma} \int d\q_{1..4} \; \frac{\prod_i^4 \e^{-\i \q_i\cdot\x_i} \, \de(\q_{1}+\q_2+\q_3+\q_4)}{\left(\mu^2+\q_1^2\right)\left(\mu^2+\q_2^2\right)\left(\mu^2+\q_3^2\right)\left(\mu^2+\q_4^2\right)},
  \label{EQ:V_Q1234}
\end{equation}
where $\alpha_\sigma=\ds\frac{ \, 3g^4\sigma}{1024 \, \pi^9 m^4}$ is a dimensionless coupling constant. To obtain this expression we used similar steps as those leading to equation (\ref{EQ:V_C1234}). The cubic $V_C$ and quartic $V_Q$ contributions to the inter-particle potential are both symmetric under 3D rotations and translations even though it is not readily evident from equations (\ref{EQ:V_C1234}) and (\ref{EQ:V_Q1234}). In Appendix C, equations (\ref{EQ:V_C1234C}) and (\ref{EQ:V_Q1234C}), we show explicitly that they indeed depend only on the inter-particle distances, \ie $V_C = V_C(x_{ij})$ and $V_Q = V_Q(x_{ij})$.

The expressions for $V_C$ and $V_Q$ can be simplified to three-dimensional quadratures:
\begin{align}
  V_C(\x_{1..4}) = & -\pi^3 \alpha_{\eta} \, \int d\x \, \Bigg\{\frac{\e^{-\mu|\x_1 + \x|}}{|\x_1 + \x|}\frac{\e^{-\mu|\x_2 + \x|}}{|\x_2 + \x|}\frac{\e^{-\mu|\x_3 + \x|}}{|\x_3 + \x|} + \frac{\e^{-\mu|\x_1 + \x|}}{|\x_1 + \x|}\frac{\e^{-\mu|\x_2 + \x|}}{|\x_2 + \x|}\frac{\e^{-\mu|\x_4 + \x|}}{|\x_4 + \x|} \nonumber \\
  & \F!\F!\F! + \frac{\e^{-\mu|\x_1 + \x|}}{|\x_1 + \x|}\frac{\e^{-\mu|\x_3 + \x|}}{|\x_3 + \x|}\frac{\e^{-\mu|\x_4 + \x|}}{|\x_4 + \x|} + \frac{\e^{-\mu|\x_2 + \x|}}{|\x_2 + \x|}\frac{\e^{-\mu|\x_3 + \x|}}{|\x_3 + \x|}\frac{\e^{-\mu|\x_4 + \x|}}{|\x_4 + \x|}\Bigg\} ,
  \label{EQ:V_C1234A} \\
  V_Q(\x_{1..4}) = & \, \pi^4 \alpha_{\sigma} \, \int d\x \; \frac{\e^{-\mu|\x_1 + \x|}}{|\x_1 + \x|}\frac{\e^{-\mu|\x_2 + \x|}}{|\x_2 + \x|}\frac{\e^{-\mu|\x_3+\x|}}{|\x_3 + \x|}\frac{\e^{-\mu|\x_4 + \x|}}{|\x_4 + \x|}.
  \label{EQ:V_Q1234A}
\end{align}
Here, analogously to the three particle case (\ref{EQ:V_C123A}), the integral expression for $V_C$, equation (\ref{EQ:V_C1234A}), is convergent for $\mu > 0$. The integrand in $V_Q$, equation (\ref{EQ:V_Q1234A}), goes as $|\x|^{-2}$ for large $|\x|$ and the integral remains finite for $\mu \ge 0$. As we shall see below, the contribution of $V_Q$ to the total inter-particle potential of the four identical particle system is particularly significant for small separations, depending on the values of the coupling constants.

The expressions for $V_C$ and $V_Q$, equations (\ref{EQ:V_C1234A}) and (\ref{EQ:V_Q1234A}), analogously to equation (\ref{EQ:V_C123B}), can be written as
\begin{align}
  V_C(\x_{1..4}, \mu) = & -\pi^3 \alpha_{\eta} \, \int d\v \, \frac{\e^{-\mu|\v|}}{|\v|} \, \Bigg\{\frac{\e^{-\mu|\v + \x_{21}|}}{|\v + \x_{21}|}\frac{\e^{-\mu|\v + \x_{31}|}}{|\v + \x_{31}|} + \frac{\e^{-\mu|\v + \x_{21}|}}{|\v + \x_{21}|}\frac{\e^{-\mu|\v + \x_{41}|}}{|\v + \x_{41}|} \nonumber \\
  & \F!\F!\F!\F!\H! + \frac{\e^{-\mu |\v + \x_{31}|}}{|\v + \x_{31}|}\frac{\e^{-\mu |\v + \x_{41}|}}{|\v + \x_{41}|} + \frac{\e^{-\mu|\v + \x_{32}|}}{|\v + \x_{32}|}\frac{\e^{-\mu|\v + \x_{42}|}}{|\v + \x_{42}|}\Bigg\},
  \label{EQ:V_C1234B} \\
  V_Q(\x_{1..4}, \mu) = & \, \pi^4 \alpha_{\sigma} \, \int d\v \; \frac{\e^{-\mu|\v|}}{|\v|}\frac{\e^{-\mu|\v + \x_{21}|}}{|\v + \x_{21}|}\frac{\e^{-\mu|\v +\x_{31}|}}{|\v + \x_{31}|}\frac{\e^{-\mu|\v + \x_{41}|}}{|\v + \x_{41}|},
  \label{EQ:V_Q1234B}
\end{align}
where $\x_{ij} = \x_i - \x_j$. 

It is not possible to obtain analytical expressions for $V_C$ and $V_Q$ in general and one must resort to numerical evaluation of these integrals. However, as for the three particle system, there is a particular solvable case, namely when $\x_1 = \x_3$ and $\x_2 = \x_4$, which shows the general features of the total potential energy. With this restriction on the coordinates there is only one inter-particle distance $x_{21} = |\x_2 - \x_1| = |\x_4 - \x_3|$ in the problem. 
%

%
\subsubsection*{Special Case: $V(\x_{1,2,1,2} ,\mu > 0) = V(x_{21}, \mu > 0)$.}
For the case when $\mu > 0$ the cubic potential term results in four copies of what was previously obtained for the three particle case, namely, 
\begin{equation}
  V_C(\x_{1,2,1,2} ,\mu > 0) = V_C(x_{21}, \mu > 0) = -\frac{8\pi^4\alpha_{\eta}}{x_{21} \mu} \Bigg\{\e^{-x_{21} \mu}\ds\left[\ln\left(3\right) - \E1\left(x_{21}\mu\right)\right] + \e^{x_{21}\mu} \, \E1\left(3x_{21}\mu\right) \Bigg\}.
  \label{EQ:V_C1212MU}
\end{equation}

The quartic potential term $V_Q(\x_{1,2,1,2}, \mu > 0)$, equation (\ref{EQ:V_Q1234B}), can be reduced to an evaluation of a single quadrature (see Appendix B for details):
\begin{align}
  V_Q(\x_{1,2,1,2}, \mu > 0) = V_Q(x_{21}, \mu > 0) = &\frac{2\pi^5\alpha_\sigma}{x_{21}} \ds\Bigg\{\int^{x_{21}}_0 \frac{dv}{v} \, \e^{-2\mu v} \bigg\{\E1\big(2\mu(x_{21}-v)\big) - \E1\big(2\mu(x_{21}+v)\big) \bigg\}\nn \\
    & \F! + \int_{x_{21}}^\infty \frac{dv}{v} \, \e^{-2\mu v} \bigg\{\E1\big(2\mu(v-x_{21})\big) - \E1\big(2\mu(x_{21}+v)\big) \bigg\}\Bigg\}.
  \label{EQ:V_Q1212MU}
\end{align}
The integrals in the expression above can not be expressed in terms of common analytic functions, thus we evaluated them numerically using the GNU Scientific Library. 

\begin{figure}[ht!]
  \begin{center}$
    \begin{array}{c}
      \includegraphics[scale = 0.6]{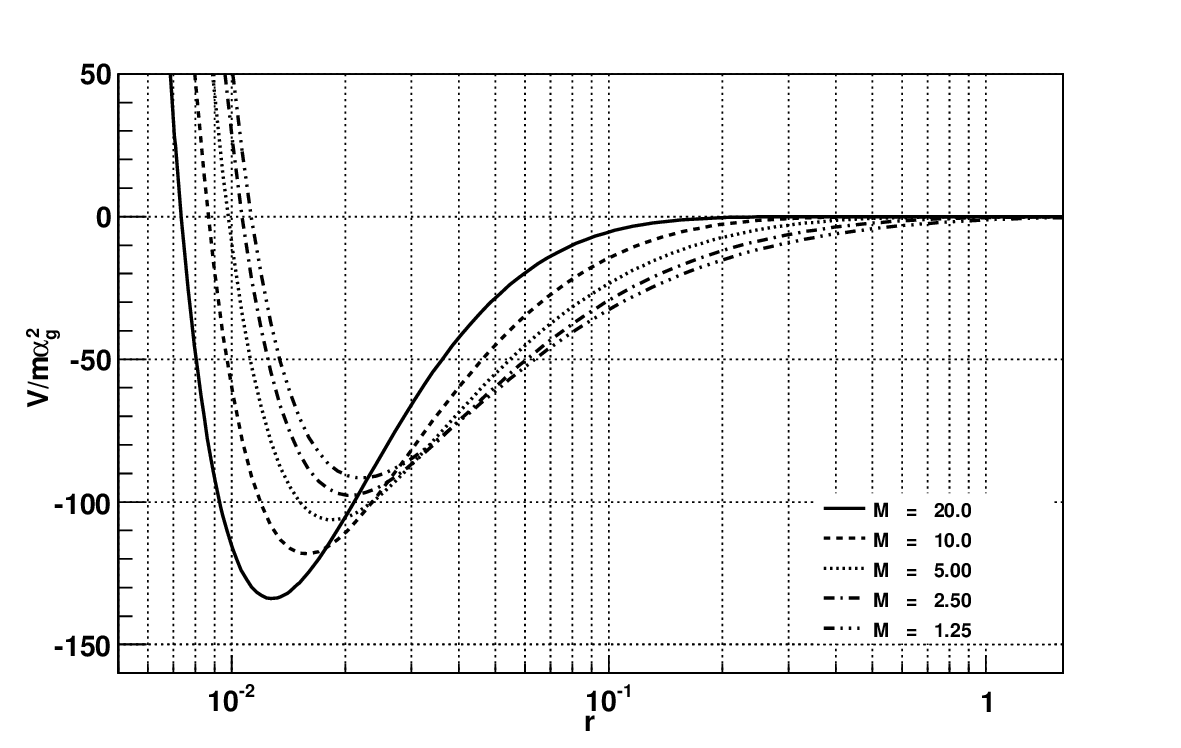}\\
      \includegraphics[scale = 0.6]{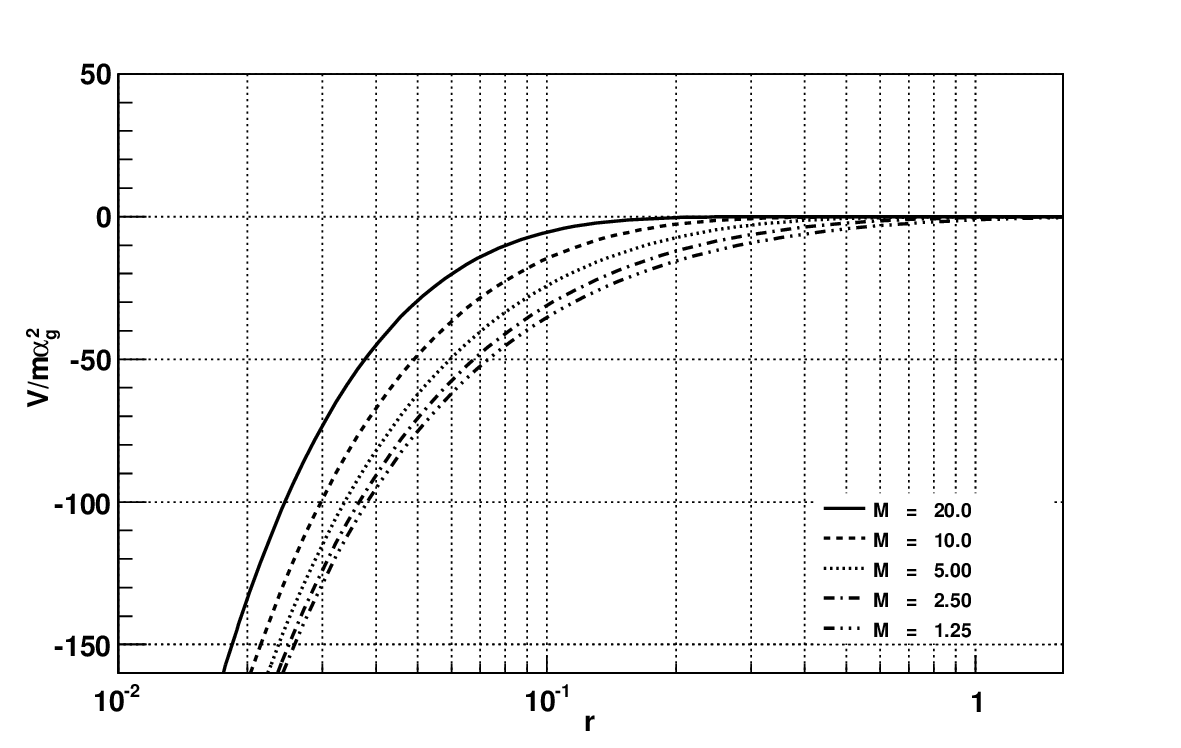}
    \end{array}$
    \caption{The inter-particle potential $V(x_{21}, \mu > 0)$ of equation (\ref{EQ:V_1212MU}) in units of $m\alpha^2_g$ as a function of $r = x_{21} \, m \alpha_g$ for the indicated values of $M$, $\kappa_1 = 0.01$, and $\kappa_2 = 0.1$ (top) and $0.005$ (bottom).}
      \label{FIG:V_1212MU}
  \end{center}
\end{figure}

The total potential energy $V = V_Y + V_C + V_Q$ in the units of $m\alpha_g^2$ as a function of $r = x_{21} \, m\alpha_g$ for the case when $\mu > 0$, suppressing two Yukawa singularities, is
\begin{align}
  V(\x_{1,2,1,2}, \mu > 0) & \; =  V(x_{12}, \mu > 0) \nn \\
   &  = m \, \alpha^2_g \, \Bigg\{ - \, 4 \, \frac{\e^{- r M}}{r} - 2 \, \kappa_1\ds\left(\ln(3) \, \frac{\e^{-rM}}{rM} - \frac{\e^{-rM}}{rM}\E1(rM) + \frac{\e^{rM}}{rM}\E1(3 \, rM)\right) \nn \\
    & \F! + \frac{\kappa_2}{r} \ds\Bigg(\int_0^{rM} \frac{dw}{w} \, \e^{- 2 w} \, \bigg\{ \E1\big(2(rM - w)\big)- \E1\big(2(rM + w)\big) \bigg\} \nn \\
    & \F!\F!\H! + \int_{rM}^\infty \frac{dw}{w} \, \e^{- 2w} \, \bigg\{ \E1\big(2(w - r M)\big) - \E1\big(2(rM + w)\big)\bigg\}\Bigg)\Bigg\} ,
  \label{EQ:V_1212MU}
\end{align}
where $w = v \, \mu $ is a dimensionless integration variable and $M = \ds\frac{\mu}{m \, \alpha_g}$ is the dimensionless mass parameter as before. Two plots of equation (\ref{EQ:V_1212MU}) are shown in Figure \ref{FIG:V_1212MU} for the parameter values $\kappa_1 = 0.01$ and  $\kappa_2 = 0.1, 0.005$. The choice $\kappa_2 = 0.1$ yields a potential which is repulsive for small separations and possesses a trough. The trough's depth increases with increasing values of the mediating field mass parameter $M$. For large separations and both values of $\kappa_2$, the potential approaches zero from below without crossing the zero of energy. 

%% file: conclusion.tex
\section*{Concluding Remarks}
The goal of this work was to derive and analyze relativistic wave equations, including the interactions kernels, for systems of three and four particles in theories that contain non-linear mediating fields. We investigated a non-linear theory (\ref{LAGRANGIAN}) in which scalar ``matter fields" $\phi$ and $\phi^*$ interact via a real non-linear scalar field $\chi$ of the Higgs type. We reformulated the theory by using the formal solution (\ref{EQ:CHI_FORMAL_SOL}) of the classical equations of motion for the field $\chi$. This enabled us to express the interaction terms of the Hamiltonian in terms of the $\chi$-field propagator and the matter fields. Since the integral equation (\ref{EQ:CHI_FORMAL_SOL}) cannot be explicitly solved for $\chi$, we used an iterative approximation and obtained the Hamiltonian density (\ref{MOD_HAMILTONIAN}), which does not contain $\chi$ explicitly but only implicitly through the $\chi$-field propagator.

We used the variational method to derive approximate few-particle eigenvalue equations for the quantized version of the theory. To this end we used Fock-space trial states, equations (\ref{TRIAL_THREE}) and (\ref{TRIAL_FOUR}), to calculate the matrix elements $\bra \Psi_3| \, {\hat H} \, | \Psi_3 \ket$ and $\bra \Psi_4| \, {\hat H} \,| \Psi_4 \ket$ for three and four identical-particle systems. Thence, we used the variational principle (\ref{EQ:VAR_PRIN}) to derive the relativistic wave equations (\ref{EQ3}) and (\ref{EQ4}) for stationary states of the three and four identical particle systems. The kernels of these integral equations represent the relativistic inter-particle interactions in momentum space. 

In order to get a picture of the inter-particle interactions in terms of familiar concepts, we considered the non-relativistic limit of the few-particle equations (Fourier transformed to co-ordinate space).

The particle-antiparticle system described by the trial state (\ref{TRIAL_TWO}) served as a preamble. For this system, the contributions of the non-linear, cubic and quartic, mediating-field terms $\eta \, \chi^3$ and $\sigma \, \chi^4$ in equation (\ref{LAGRANGIAN}) are not sampled by the trial state (\ref{TRIAL_TWO}), and so the non-relativistic particle-antiparticle potential is just the Yukawa plus virtual-annihilation contact potential (\ref{PE2}). To probe the effect of the non-linear terms, one would need to consider a more elaborate trial state than (\ref{TRIAL_TWO}), such as one containing more Fock-space components, for example 
\begin{align}
  |\widetilde{\Psi}_2\ket & = | \Psi_2 \ket + | \Psi_4 \ket \nn \\ 
  & = \ds \int d\p_{1,2} \; F({\p_{1,2}) \; a^{\dagger}(\p_1) \, b^{\dagger}(\p_2) |0\ket + \ds \int d\p_{1..4} \; G(\p_{1..4}}) \; a^{\dagger}(\p_1) \, b^{\dagger}(\p_2) \, a^{\dagger}(\p_3) \,b^{\dagger}(\p_4) |0\ket.
  \label{TRIAL_IMPROVED}
\end{align}
The trial state (\ref{TRIAL_IMPROVED}) leads to coupled multidimensional integral equations for the coefficient functions $F$ and $G$. These equations would need to be solved approximately (say, variationally), which is a tedious, if straightforward task. The extraction of information about the particle-antiparticle potential would be an additional challenge. 

Reformulation of the theory enabled us to use simple three and four particle trial states, equations (\ref{TRIAL_THREE}) and (\ref{TRIAL_FOUR}), to derive relativistic wave equations and interaction kernels which include novel additional terms due to the non-linear terms of the Hamiltonian density $\cH_{I_2}$ and $\cH_{I_3}$. The non-relativistic limits of these interactions, in coordinate space, are three and four-point inter-particle potentials which are shown to depend on the inter-particle distances only.

Unfortunately, the potentials (\ref{EQ:V_C123A}), (\ref{EQ:V_C1234B}) and (\ref{EQ:V_Q1234B}) due to the non-linear interaction terms cannot be evaluated analytically for arbitrary inter-particle distances, and hence must be evaluated numerically. However, analytic expressions are obtainable for particular, restricted cases, for arbitrary values of the mass $\mu$ of the mediating field quantum, and these provide a good picture of the general form of the potentials.

For the three particle system, the analytic expressions of the inter-particle potentials with certain coordinate restrictions are given in equations (\ref{EQ:V_122MU}) and (\ref{EQ:V_DMU}) with $\mu > 0$. We find that the Yukawa potential is augmented by the cubic term $V_C$ as illustrated in Figures \ref{FIG:V_122SEPARATE}, \ref{FIG:V_122MU} and \ref{FIG:V_DMU}. The potential does not rise above the reference line of the rest mass (\ie $2m$) as the inter-particle separation increases but simply approaches zero. This indicates that the potential does not support bound states with energies greater than the rest mass of the constituents. 

For the three particle system, the numerical evaluation of the potential in the case of arbitrary inter-particle distances is performed where one of the three inter-particle distances is kept constant. These surface plots are displayed in Figure \ref{FIG:V_surface}. Similarly, they show that the inter-particle potential does not support bound states with energies larger than the sum of the masses of the constituent particles.

For the four particle trial state with $\mu > 0$, the analytic expression of the potential for the particular case $\x_1 = \x_3$ and $\x_2 = \x_4$ is given in equation (\ref{EQ:V_1212MU}). The quartic term can alter the behaviour of the potential for small values of the inter-particle separation depending on the coupling constants, as is shown in Figure \ref{FIG:V_1212MU}. High values of the mass parameter $M$ are seen to be responsible for the development of a trough in the potential, provided that the coupling constants $\kappa_1$ and $\kappa_2$ are such that the potential is repulsive for small $r$. Unfortunately, it is not possible to give graphical representations of the potentials for arbitrary values of the six $x_{ij}$ coordinates.

In this work we have concentrated on the nature of the inter-particle interactions in a scalar model with a non-linear Higgs-type mediating field. The non-relativistic limit was considered in some detail. We shall leave for the future the challenging task of solving the three and four particle relativistic and non-relativistic equations derived in this paper, as well as the analysis of the particle-antiparticle system with more elaborate trial states. 

%% file: appendixA.tex
\section*{Appendix A: Three Particle State}
In this section we present some intermediate steps of the derivations with the three particle trial state (\ref{TRIAL_THREE}). All matrix elements were calculated using Maple. 
\subsection*{Derivation of the kernels}
The matrix element in the Schr\"odinger picture for the three identical particle trial state (\ref{TRIAL_THREE}) is given by 
\begin{equation}
  \bra \Psi_3 | \, \hat{H} - E \, | \Psi_3 \ket =   \bra \Psi_3 | \, \hH_{\phi} + \hH_{I_1} + \hH_{I_2} - E \, | \Psi_3 \ket 
\end{equation}
where the contributions are
\begin{align}
  \bra \Psi_3 | \, \hat{H}_{\phi} -E \, | \Psi_3 \ket = & \int d\p_{1,2,3} \big(\omega_{\p_1} + \omega_{\p_2} + \omega_{\p_3} - E \big) \; F^{\ast}(\p^{\prime}_{1,2,3}) \, F_S(\p_{1,2,3}), \\
  \bra \Psi_3 | \, \hat{H}_{I_{1}} \, | \Psi_3 \ket =  & -\frac{g^2}{8\,(2\pi)^{3}} \int \frac{d\p^{\prime}_{1,2,3} \, d\p_{1,2,3}}{\ds\sqrt{\omega_{\p^{\prime}_{1}}\omega_{\p^{\prime}_{2}}\omega_{\p_1}\omega_{\p_2}}} \; F^{\ast}_S(\p^{\prime}_{1,2,3}) \, F_S(\p_{1,2,3}) \nonumber \\
  & \times \delta(\p^{\prime}_1 + \p^{\prime}_2 - \p_1 - \p_2) \, \delta(\p^{\prime}_3 - \p_3) \, \left[\frac{1}{\mu^2-(p^{\prime}_1 - p_1)^2}\right], \\
  \bra \Psi_3 | \, \hat{H}_{I_{2}} \, | \Psi_3 \ket = & - \frac{g^{3}\eta}{8(2\pi)^{6}} \int \frac{d\p^{\prime}_{1,2,3} \, d\p_{1,2,3}}{\ds\sqrt{\omega_{\p^{\prime}_{1}}\omega_{\p^{\prime}_{2}}\omega_{\p^{\prime}_{3}}\omega_{\p_1}\omega_{\p_2}\omega_{\p_3}}} \; F^{\ast}_S(\p^{\prime}_{1,2,3}) \, F_S(\p_{1,2,3}) \nonumber \\
  & \times \delta(\p^{\prime}_1 + \p^{\prime}_2 +\p^{\prime}_3 - \p_1 - \p_2 - \p_3) \nonumber \\
  & \times \left[\frac{1}{\mu^2-(p^{\prime}_1 + p^{\prime}_2 - p_1 - p_2)^2}\frac{1}{\mu^2-(p^{\prime}_1 - p_1)^2}\frac{1}{\mu^2-(p^{\prime}_2 - p_2)^2}\right]. 
\end{align}
In working out the variational derivative the following identity was used:
\begin{equation}
  \frac{\delta F_{S}(\p_{1,2,3})}{\delta F(\q_{1,2,3})} = \sum_{i_1, i_2, i_3}^6 \delta(\p_{i_1} - \q_1) \, \delta(\p_{i_2} - \q_2) \, \delta(\p_{i_3} - \q_3),
\end{equation}
where the summation is on the six permutation of the indices $i_1, i_2$ and $i_3$. 

The relativistic kernels for the three particle trial state are given in equations (\ref{EQ:Y_33}) and (\ref{EQ:C_33}) in the body of the paper. In the non-relativistic limit the kernels become
\begin{align}
  Y_{3,3}(\p^{\prime}_{1,2,3}, \p_{1,2,3}) =  & -\frac{g^2}{8\,(2\pi)^{3}m^2} \; \left[\frac{\delta(\p^{\prime}_1 + \p^{\prime}_2 - \p_1 - \p_2) \, \delta(\p^{\prime}_3 - \p_3)}{\mu^2+(\p^{\prime}_1 - \p_1)^2}\right. \nonumber \\
    & + \left. \frac{\delta(\p^{\prime}_1 + \p^{\prime}_3 - \p_1 - \p_3) \, \delta(\p^{\prime}_2 - \p_2)}{\mu^2+(\p^{\prime}_3 - \p_3)^2} +  \frac{\delta(\p^{\prime}_2 + \p^{\prime}_3 - \p_2 - \p_3) \, \delta(\p^{\prime}_1 - \p_1)}{\mu^2+(\p^{\prime}_2 - \p_2)^2}\right] \\
  C_{3,3}(\p^{\prime}_{1,2,3}, \p_{1,2,3}) = & - \frac{3 \, g^3\eta}{4 \, (2\pi)^6m^3} \, \frac{\delta(\p^{\prime}_1 + \p^{\prime}_2 +\p^{\prime}_3 - \p_1 - \p_2 - \p_3)}{(\mu^2+(\p^{\prime}_1 - \p_1)^2)(\mu^2+(\p^{\prime}_2 - \p_2)^2)(\mu^2+(\p^{\prime}_3 - \p_3)^2)}
  \label{EQ:C33AA}
\end{align}
where we used the symmetry property of $F_S$ to simplify the expression. 

To obtain equation (\ref{EQ:V_C123}) we multiplied equation (\ref{EQ:C33AA}) by $\ds\prod_i^3\, \e^{\i\p_i\cdot\x_i}$ and shifted the variable of integration $\p^{\prime}_i - \p_i = \q_i$. Integrating over the vectors $\q_i$ leads to equation (\ref{EQ:V_C123A}). This integration is performed using the standard technique where the radial integral is evaluated in the complex plane using Cauchy's integration formula. 
\subsection*{Derivation of the cubic potential energy $V_C$ for $\x_2 = \x_3$}
The cubic inter-particle potential for the three particle trial state (\ref{TRIAL_THREE}) in the case when $\mu > 0$ and $\x_2 = \x_3$ is determined from the following expression:  
\begin{equation}
  V_C(\x_{1,2,2}, \mu) = V_C(x_{21}, \mu)= - \alpha_{\eta}\pi^3 \, \int d\v \, \frac{\e^{-\mu v}}{v} \, \frac{\e^{-2\mu|\v+\x_{21}|}}{|\v+\x_{21}|^2},
  \label{EQ:STEP1}
\end{equation}
where  $\x_{21} = \x_2 - \x_1$ and the non-vector notation is understood to mean the magnitude of the vector. After the trivial azimuthal integration we obtain: 
\begin{equation}
  V_C(x_{21}, \mu) = - 2 \pi^4\alpha_{\eta} \, \int_0^\infty dv \, v^2 \int_{-1}^{+1} \, dw \; \frac{\e^{-2 \, \mu v}}{v^2}\frac{\e^{-\mu|\v+\x_{21}|}}{|\v+x_{21}|}, 
  \label{EQ:STEP1A}
\end{equation}
where $w$ = $\cos\theta$ with $\theta$ being the polar angle of the vector $\v$. The polar integration is performed using the substitution $\varrho^2=x_{21}^2+v^2+2 \,x_{21} \, v \, w$, with $\varrho_{1}=(x_{21}+v)$ and $\varrho_{2}=|x_{21}-v|$ as the upper and the lower limits of integration. Thereupon, the integral of equation (\ref{EQ:STEP1A}) can be written as
\begin{equation}
  V_C(x_{21}, \mu) = -\frac{2\pi^4\alpha_{\eta}}{x_{21}} \, \int_0^{\infty} \frac{dv}{v} \, \e^{-2\mu v} \int_{\vr_2}^{\vr_1} d\vr \, \e^{-\mu \vr}.
  \label{EQ:STEP2} 
\end{equation}
We integrate over the variable $\vr$, split the interval of integration accordingly and end up with the following radial integral: 
\begin{equation}
  V_C(x_{21}, \mu) = - \frac{2\pi^4\alpha_{\eta}}{\mu x_{21}} \, \Bigg\{\e^{-\mu x_{21}} \int_0^{x_{21}} \frac{dv}{v} \, \left(\e^{-\mu v}- \e^{-3 \mu v}\right) + \left(\e^{\mu x_{21}} - \e^{-\mu x_{21}} \right) \int_{x_{21}}^{\infty} \frac{dv}{v} \, \e^{-3\mu v} \Bigg\}.
\end{equation}
The integral is expressible in terms of the exponential integral defined by equation (\ref{EQ:EXPINT}). Thus, the cubic potential term for the three particle trial state in the case when $\mu > 0$ and $\x_2 = \x_3$ evaluates to
\begin{equation}
  V_C(x_{21}, \mu) = -\frac{2\pi^4\alpha_{\eta}}{x_{21} \mu} \Bigg\{\e^{-x_{21} \mu}\ds\left[\ln\left(3\right) - \E1\left(x_{21} \mu\right)\right] + \e^{x_{21} \mu} \, \E1\left(3x_{21} \mu\right) \Bigg\},
\end{equation}
which is equation (\ref{EQ:V_C122MU}).

%% file: appendixB.tex
\section*{Appendix B: Four Particle State}
In this section we present some intermediate steps of the derivations with the four particle trial state (\ref{TRIAL_FOUR}). All matrix elements were calculated using Maple. 
\subsection*{Derivation of the kernels}
The matrix element in the Schr\"odinger picture for the four identical particle trial state (\ref{TRIAL_FOUR}) is given by 
\begin{equation}
  \bra \Psi_4 | \, \hat{H} - E \, | \Psi_4 \ket =   \bra \Psi_4 | \, \hH_{\phi} + \hH_{I_1} + \hH_{I_2} + \hH_{I_3} - E \, | \Psi_4 \ket, 
\end{equation}
where the contributions are
\begin{align}
  \bra\Psi_4| \, \hat{H}_{\phi} - E \, |\Psi_4 \ket  = &  \int d\p_{1..4} \; F^{\ast}(\p_{1..4}) \, F_{S}(\p_{1..4}) \; \ds\left(\omega_{\p_1} + \omega_{\p_2} + \omega_{\p_3} + \omega_{\p_4} - E \right), \nn \\
  \bra \Psi_4| \, \hat{H}_{I_1} \, | \Psi_4 \ket = & - \frac{g^2}{16(2\pi)^3} \; \int \frac{d\p^{\prime}_{1..4} \, d\p_{1..4}}{\ds\sqrt{\omega_{\p^{\prime}_1}\omega_{\p^{\prime}_2}\omega_{\p_1}\omega_{\p_2}}} \, F^{\ast}_{S}(\p^{\prime}_{1..4}) \, F_{S}(\p_{1..4}) \nn \\
  & \times \, \de(\p^{\prime}_1 + \p^{\prime}_2 - \p_1 - \p_2)\, \de(\p^{\prime}_3 - \p_3) \, \de(\p^{\prime}_4 - \p_4) \, \left[\frac{1}{\mu^2-(p^{\prime}_{1}-p_{1})^2}\right], \\
  \bra\Psi_4 | \,\hat{H}_{I_2} \, | \Psi_4 \ket = & - \frac{g^3\eta}{8(2\pi)^6} \, \int  \frac{d\p^{\prime}_{1..4} \, d\p_{1..4}}{\ds\sqrt{\omega_{\p^{\prime}_1}\omega_{\p^{\prime}_{2}}\omega_{\p^{\prime}_3}\omega_{\p_1}\omega_{\p_2}\omega_{\p_3}}} \, F^{\ast}_S(\p^{\prime}_{1..4}) \, F_S(\p_{1..4})   \nonumber \\
  & \times \, \de(\p^{\prime}_1 + \p^{\prime}_2 + \p^{\prime}_3 - \p_1 - \p_2 - \p_3)\,  \de(\p^{\prime}_4 - \p_4) \nonumber \\
  & \times \, \left[\frac{1}{\mu^2-(p^{\prime}_1 + p^{\prime}_2 - p_1 - p_2)^2}\frac{1}{\mu^2-(p^{\prime}_1 - p_1)^2}\frac{1}{\mu^2-(p^{\prime}_2 - p_2)^2}\right],
\end{align}
and
\begin{align}
  \bra \Psi_4 | \, \hat{H}_{I_{3}} \, | \Psi_4 \ket = \, & \frac{g^4\sigma}{16(2\pi)^9} \; \int \frac{d\p^{\prime}_{1..4} \, d\p_{1..4}}{\ds\sqrt{\omega_{\p^{\prime}_1}\omega_{\p^{\prime}_2}\omega_{\p^{\prime}_3}\omega_{\p^{\prime}_4}\omega_{\p_1}\omega_{\p_2}\omega_{\p_3}\omega_{\p_4}}} \, F^{\ast}_{S}(\p^{\prime}_{1..4}) \, F_{S}(\p_{1..4}) \nn \\
  & \times \, \de(\p^{\prime}_1 + \p^{\prime}_2 + \p^{\prime}_3 + \p^{\prime}_4 - \p_1 - \p_2 - \p_3 - \p_4) \nn \\ 
  & \times \, \left[\frac{1}{\mu^2-(p^{\prime}_1 + p^{\prime}_2 + p^{\prime}_3 - p_1 - p_2 - p_3)^2}\right]  \nn \\
  & \times \, \left[\frac{1}{\mu^2-(p^{\prime}_1 - p_1)}\frac{1}{\mu^2-(p^{\prime}_2 - p_2)^2}\frac{1}{\mu^2-(p^{\prime}_3 - p_3)^2}\right].
\end{align}

The following identity was used in working out the variational derivative:
\begin{equation}
  \frac{\delta F_{S}(\p_{1..4})}{\delta F(\q_{1..4})} = \sum_{i_1, i_2, i_3, i_4}^{24} \de(\p_{i_1} - \q_1) \, \delta(\p_{i_2} - \q_2) \, \delta(\p_{i_3} - \q_3) \,  \delta(\p_{i_4} - \q_4)
\end{equation}
where the summation is on the 24 permutation of the indices $i_1, i_2, i_3$ and $i_4$. 

The relativistic Yukawa and cubic interaction kernels for the the four particle trial state (\ref{TRIAL_FOUR}) are
\begin{align}
  \cY_{4,4} (\p^{\prime}_{1..4}, \p_{1..4}) = & -\frac{g^2}{16(2\pi)^3} \nn \\
  & \H! \times \sum_{i_1, i_2, i_3, i_4}^{24} \frac{\de(\p^{\prime}_{1}+\p^{\prime}_{2}-\p_{i_1}-\p_{i_2})\, \de(\p^{\prime}_{3}-\p_{i_3}) \, \de(\p^{\prime}_{4}-\p_{i_4})}{\ds\sqrt{\omega_{\p^{\prime}_1} \omega_{\p^{\prime}_{2}} \omega_{\p_{i_1}}\omega_{\p_{i_2}}}} \left[\frac{1}{\mu^2-(p^{\prime}_{1}-p_{i_1})^2}\right], \\
  \cC_{4,4} (\p^{\prime}_{1..4}, \p_{1..4}) = & - \frac{g^3\eta}{8(2\pi)^6} \, \sum^{24}_{i_1, i_2, i_3,i_4} \frac{ \de(\p^{\prime}_1+\p^{\prime}_2+\p^{\prime}_3-\p_{i_1}-\p_{i_2}-\p_{i_3}) \,\de(\p^{\prime}_4-\p_{i_4})}{\ds\sqrt{\omega_{\p^{\prime}_1}\omega_{\p^{\prime}_2}\omega_{\p^{\prime}_3}\omega_{\p_{i_1}}\omega_{\p_{i_2}}\omega_{\p_{i_3}}}} \nn \\
  & \H! \times \, \left[\frac{1}{\mu^2-(p^{\prime}_1+p^{\prime}_2-p_{i_1}-p_{i_2})^2}\frac{1}{\mu^2-(p^{\prime}_1-p_{i_1})^2}\frac{1}{\mu^2-(p^{\prime}_2-p_{i_2})^2}\right].
\end{align}
The relativistic quartic kernel for the four particle trial state is given in equation (\ref{EQ:Q44}).
In the non-relativistic limit the interaction kernels for the four particle trial state reduce to
\begin{align}
  Y & (\p^{\prime}_{1..4}, \p_{1..4}) = \frac{g^2}{4(2\pi)^3m^2} \nn \\
  & \times \, \ds\left[\frac{\de(\p^{\prime}_1 + \p^{\prime}_2 - \p_1 - \p_2) \, \de(\p^{\prime}_3 - \p_3) \, \de(\p^{\prime}_4 - \p_4)}{\mu^2+(\p^{\prime}_1-\p_1)^2} + \frac{\de(\p^{\prime}_1 + \p^{\prime}_3 - \p_1 - \p_3) \, \de(\p^{\prime}_2 - \p_2) \, \de(\p^{\prime}_4 - \p_4)}{\mu^2+(\p^{\prime}_1-\p_1)^2}\right. \nn\\
    & \H! + \frac{\de(\p^{\prime}_1 + \p^{\prime}_4 - \p_1 -\p_4) \, \de(\p^{\prime}_2 - \p_2) \, \de(\p^{\prime}_3 - \p_3)}{\mu^2+(\p^{\prime}_1 - \p_1)^2}+\frac{\de(\p^{\prime}_2 + \p^{\prime}_3 - \p_2 - \p_3) \, \de(\p^{\prime}_1 - \p_1) \, \de(\p^{\prime}_4 - \p_4)}{\mu^2+(\p^{\prime}_2 - \p_2)^2} \nn \\
    & \H! + \left.\frac{\de(\p^{\prime}_2 + \p^{\prime}_4 - \p_2 - \p_4) \, \de(\p^{\prime}_1 - \p_1) \, \de(\p^{\prime}_3 - \p_3)}{\mu^2+(\p^{\prime}_2-\p_2)^2} + \frac{\de(\p^{\prime}_3 + \p^{\prime}_4 - \p_3 - \p_4) \, \de(\p^{\prime}_1 - \p_1) \, \de(\p^{\prime}_2 - \p_2)}{\mu^2 + (\p^{\prime}_3- \p_3)^2}\right], \\
  C & (\p^{\prime}_{1..4}, \p_{1..4}) = \frac{3 \, g^3\eta}{4 \, (2\pi)^6m^3} \nn \\
  & \F!\F!\F!\H! \times \ds \left[\frac{\delta(\p^{\prime}_1 + \p^{\prime}_2 + \p^{\prime}_3 - \p_1 - \p_2 - \p_3) \, \de(\p^{\prime}_4 - \p_4)}{(\mu^2+(\p^{\prime}_1-\p_1)^2)(\mu^2+(\p^{\prime}_2 - \p_2)^2)(\mu^2+(\p^{\prime}_3 - \p_3)^2)}\right. \nn \\
    & \F!\F!\F!\F! + \frac{\de(\p^{\prime}_1 + \p^{\prime}_2 + \p^{\prime}_4 - \p_1 - \p_2 - \p_4) \, \de(\p^{\prime}_3 - \p_3)}{(\mu^2+(\p^{\prime}_1 - \p_1)^2)(\mu^2+(\p^{\prime}_2 - \p_2)^2)(\mu^2+(\p^{\prime}_4 - \p_4)^2)} \nn \\
    & \F!\F!\F!\F! + \frac{\de(\p^{\prime}_1 + \p^{\prime}_3 + \p^{\prime}_4 - \p_1 - \p_3 - \p_4) \, \de(\p^{\prime}_2 - \p_2)}{(\mu^2+(\p^{\prime}_1 - \p_1)^2)(\mu^2+(\p^{\prime}_3 - \p_3)^2)(\mu^2 + (\p^{\prime}_4 - \p_4)^2)} \nn \\
    & \F!\F!\F!\F! + \left.\frac{\de(\p^{\prime}_2 + \p^{\prime}_3 + \p^{\prime}_4 - \p_2 - \p_3 - \p_4) \, \de(\p^{\prime}_1 - \p_1)}{(\mu^2+(\p^{\prime}_2 - \p_2)^2)(\mu^2+(\p^{\prime}_3 -\p_3)^2)(\mu^2+(\p^{\prime}_4 - \p_4)^2)}\right], 
\end{align}
and
\begin{align}
  Q & (\p^{\prime}_{1..4}, \p_{1..4})  = - \frac{3 \, g^4\sigma}{2(2\pi)^9m^4} \nn \\
  & \F!\F!\F!\H! \times \left[\frac{\de(\p^{\prime}_1 + \p^{\prime}_2 + \p^{\prime}_3 + \p^{\prime}_4 - \p_1 - \p_2 - \p_3 - \p_4)}{(\mu^2+(\p^{\prime}_1-\p_1)^2)(\mu^2+(\p^{\prime}_2-\p_2)^2)(\mu^2+(\p^{\prime}_3-\p_3)^2)(\mu^2+(\p^{\prime}_4-\p_4)^2)}\right].
\end{align}
\subsection*{Derivation of the quartic potential energy $V_Q$ for $\x_1 = \x_3$ and $\x_2 = \x_4$}
The quartic inter-particle potential term for the four particle trial state (\ref{TRIAL_FOUR}) in the case when $\mu > 0$, $\x_1 = \x_3$ and $\x_2 = \x_4$ is determined from the following expression:
\begin{equation}
  V_Q(\x_{1,2,1,2}, \mu > 0) = V_Q(x_{21}, \mu > 0) = \alpha_{\sigma}\pi^4 \, \int d\v \; \frac{\e^{-2 \, \mu |\v|}}{|\v|^2}\frac{\e^{- 2 \, \mu |\v + \x_{21}|}}{|\v + \x_{21}|^2}.
\end{equation}
Following similar steps as those in equations (\ref{EQ:STEP1})-(\ref{EQ:STEP2}) leads to 
\begin{equation}
  V_Q(\x_{1,2,1,2}, \mu > 0) = V_Q(x_{21}, \mu > 0) = \frac{2\pi^5\sigma}{x_{21}} \int \frac{dv}{v} \, \e^{-2\mu v}\int^{\varrho_1}_{\varrho_2} \frac{d\varrho}{\varrho} \, \e^{- 2\mu \varrho},
\end{equation}
where $\varrho_1$, $\varrho_2$ and $x_{21}$ are given below equation (\ref{EQ:STEP1A}). The result of the integration over the variable $\varrho$ can be expressed in terms of the exponential integral (\ref{EQ:EXPINT}) and is given in equation (\ref{EQ:V_Q1212MU}). The remaining integrals have to be evaluated numerically.

%% file: appendixC.tex
\section*{Appendix C: Gaussian Parametrization}
In this section we present an alternative method for evaluating the cubic and the quartic potential contributions $V_C$ and $V_Q$ for the three and four identical particle cases. The invariance of the potentials $V_C$ and $V_Q$ under rotations and translations of the coordinates becomes explicit in this method. 

We shall write the denominators of equations (\ref{EQ:V_C123}), (\ref{EQ:V_C1234}) and (\ref{EQ:V_Q1234}) using a technique which enables us to perform Gaussian integration over the momentum variables. We use the following identity and the Gaussian integration formula:
\begin{align}
  \int_{0}^{\infty}d\beta \, \e^{-A\beta} = \, & \frac{1}{A} \H! \tx{for} \H! A > 0, 
  \label {EQ:WILSON} \\
  \int_{-\infty}^{\infty}\prod_{i} \, dx_{i} \, \exp\left[-\frac{1}{2}K_{ij}x_{i}x_{j}-L_{j}x_{j}-W\right] = \, & \textrm{Det} \, \left[\frac{K}{2\pi}\right]^{-\frac{1}{2}}\exp\left[\frac{1}{2}K^{-1}_{ij}L_{i}L_{j}-W\right],
  \label{EQ:GAUSS}
\end{align}
\ni where $K$ is an invertible and symmetric matrix, $L$ is a vector and $W$ is a constant. 
\subsection*{Three Particle State}
The three particle trial state (\ref{TRIAL_THREE}) yields the cubic interaction kernel (\ref{EQ:V_C123}). After integrating over the delta function in (\ref{EQ:V_C123}) there remain only two momentum integrations. Using the identity of equation (\ref{EQ:WILSON}) on each factor separately, we obtain the following expression:
\begin{equation}
  V_C(\x_{ij}, \mu > 0) =  - \pi^3 \alpha_{\eta} \int_{0}^{\infty} d\beta_{1,2,3} \int d\q_{1,2} \; \e^{ -(\mu^2 + \q^2_1) \, \beta_1 - (\mu^2 + \q^2_2) \, \beta_2 - (\mu^2 + (\q_1 +\q_2)^2) \, \beta_3} \e^{-\i\q_1\cdot\x_{21}} \, \e^{-\i\q_2\cdot\x_{31}},
  \label{EQ:V_C123M2A}
\end{equation}
where $\x_{ij} = \x_i - \x_j$ are the inter-particle vectors. Upon expanding the squares in the exponentials and defining $\q=\q_1+\q_1$  we identify a six dimensional Gaussian integral. The matrix $K$ and its inverse $K^{-1}$, the vectors $L_{i}$ and $W$ for the Gaussian integral are given in block diagonal form
\begin{align}
  K = & \left[
    \begin{array}{cc}
      2(\beta_{1}+\beta_{3}) & 2\beta_{3}  \\
      2\beta_{3} & 2(\beta_{2}+\beta_{3})  \\ \end{array}
    \right] , \;  
  K^{-1}=\frac{1}{4(\beta_{1}\beta_{2}+\beta_{1}\beta_{3}+\beta_{2}\beta_{3})}\left[
    \begin{array}{cc}
      2(\beta_{2}+\beta_{3}) & -2\beta_{3}  \\
      -2\beta_{3} & 2(\beta_{1}+\beta_{3})  \\ \end{array}
    \right] 
  \label{EQ:MK3}  \\
  L= & \i\left[
    \begin{array}{cc}
      \x_{21}, & \x_{31} \end{array}
    \right],\F!
  W = \left(\beta_{1}+\beta_{2}+\beta_{3}\right)\mu^{2}.
  \label{EQ:VLM}
\end{align}
Applying equation (\ref{EQ:GAUSS}) we arrive at
\begin{equation}
  V_C(\x_{ij}, \mu > 0) = - \pi^3 \alpha_{\eta} \int_{0}^{\infty}d\beta_{1,2,3} \; \frac{\e^{-\mu^{2}(\beta_{1}+\beta_{2}+\beta_{3})}}{(\beta_{1}\beta_{2}+\beta_{1}\beta_{3}+\beta_{2}\beta_{3})^{3/2}} \; \exp\left[-\frac{\beta_{1}\x_{21}^{2}+\beta_{2}\x_{31}^{2}+\beta_{3}\x_{32}^{2}}	{4\left(\beta_{1}\beta_{2}+\beta_{1}\beta_{3}+\beta_{2}\beta_{3}\right)}\right]. 
\end{equation}
This expression shows explicitly that the cubic potential term $V_C$ depends only on the inter-particle distances. Numerical integration over the parameters $\beta_i$ is required to complete the calculation.
\subsection*{Four Particle State}
The four particle trial state (\ref{TRIAL_FOUR}) yields the cubic interaction kernel (\ref{EQ:V_C1234}). The calculation of $V_C$ for the four particle trial state is identical to that of the three particle trial state. There are basically four copies of the three particle results with different inter-particle distances involved. The matrix $K$ and its inverse $K^{-1}$ are identical for all four terms and the same as for the three particle trial state equation (\ref{EQ:MK3}). The vectors $L_i$ pertaining to each term in equation (\ref{EQ:V_C1234}) are
\begin{equation}
  L_1 = \i\left[
    \begin{array}{cc}
      \x_{21}, & \x_{31} \end{array}
    \right],\,
  L_2 = \i\left[
    \begin{array}{cc}
      \x_{21}, & \x_{41} \end{array}
    \right],\,
  L_3 = \i\left[
    \begin{array}{cc}
      \x_{31}, & \x_{41} \end{array}
    \right],\,
  L_4 = \i\left[
    \begin{array}{cc}
      \x_{32}, & \x_{42} \end{array}
    \right]
\end{equation}
where the subscript indicates the corresponding term in equation (\ref{EQ:V_C1234}). The constant $W$ is the same as in equation (\ref{EQ:VLM}). Application of the Gaussian integration formula (\ref{EQ:GAUSS}) leads to the result
\begin{align}
  V_C(\x_{ij}\, \mu > 0) = & -\alpha_{\eta} \pi^3 \, \int^{\infty}_0 d\b_{1,2,3} \, \frac{\e^{-\mu^2(\b_1+\b_2+\b_3)}}{(\b_1\b_2+\b_1\b_3+\b_2\b_3)^{3/2}} \nonumber \\
  & \times \, \Bigg\{\exp\left(-\frac{\b_1\x^2_{12}+\b_2\x^2_{13}+\b_3\x^2_{23}}{4(\b_1\b_2+\b_1\b_3+\b_2\b_3)}\right) + \exp\left(-\frac{\b_1\x^2_{12}+\b_2\x^2_{14}+\b_3\x^2_{24}}{4(\b_1\b_2+\b_1\b_3+\b_2\b_3)}\right) \nonumber \\
  & \F!\; \exp\left(-\frac{\b_1\x^2_{13}+\b_2\x^2_{14}+\b_3\x^2_{34}}{4(\b_1\b_2+\b_1\b_3+\b_2\b_3)}\right) + \exp\left(-\frac{\b_1\x^2_{23}+\b_2\x^2_{24}+\b_3\x^2_{34}}{4(\b_1\b_2+\b_1\b_3+\b_2\b_3)}\right)\Bigg\}.
  \label{EQ:V_C1234C}
\end{align}
From equation (\ref{EQ:V_C1234C}) it is evident that $V_C$ depends on the inter-particle distances $x_{ij} = |\x_i - \x_j|$ only.

The four particle trial state yields the quartic interaction kernel equation (\ref{EQ:V_Q1234}). The calculation follows the same steps as for the cubic interaction kernel. Applying the identity of equation (\ref{EQ:WILSON}) leads to a 9 dimensional Gaussian integral. The matrix $K$ and its inverse $K^{-1}$ in block diagonal form are
\begin{align}
  K = &\left[
    \begin{array}{ccc}
      2(\b_1+\b_4) & 2\b_4         & 2\b_4   \\
      2\b_4        & 2(\b_2+\b_4)  & 2\b_4   \\ 
      2\b_4        & 2\b_4         & 2(\b_3+\b_4) 
    \end{array}
    \right], 
  K^{-1}=\frac{1}{2\b_{1234}}\left[
    \begin{array}{ccc}
      \b_{234}     & -\b_{3}\b_{4} & -\b_{2}\b_{4}  \\
      -\b_3\b_4    & \b_{134}      & -\b_1\b_4      \\
      -\b_2\b_4    & -\b_1\b_4     & \b_{124}      
    \end{array}
    \right],
\end{align}
where $\b_{1234}=\b_1\b_2\b_3+\b_1\b_3\b_4+\b_1\b_2\b_4+\b_2\b_3\b_4$ and $\b_{ijk}=\b_{ij}+\b_{ik}+\b_{jk}$. The vector $L$ and the constant $W$ of the Gaussian integration are 
\begin{equation}
  L=\i \left(\x_{41}, \, \x_{42}, \, \x_{43}\right), \F! W = \left(\b_1+\b_2+\b_3+\b_4\right)\mu^{2}.
\end{equation}
Applying the Gaussian integration formula (\ref{EQ:GAUSS}) and after some algebra we end up with the expression
\begin{multline}
  V_Q(\x_{ij}, \mu > 0) = \alpha_{\sigma} \pi^{9/2} \, \int^{\infty}_{0} d\b_{1..4} \, \frac{\e^{-\mu^2(\b_1+\b_2+\b_3+\b_4)}}{(\b_{1234})^{3/2}} \\
  \times \, \exp\left(-\frac{\b_3\b_4\x^2_{12}+\b_2\b_4\x^2_{13}+\b_2\b_3\x^2_{14}+\b_1\b_4\x^2_{23}+\b_1\b_3\x^2_{24}+\b_1\b_2\x^2_{34}}{4\b_{1234}}\right).
  \label{EQ:V_Q1234C}
\end{multline}
This expression shows explicitly that the quartic potential term $V_Q$ depends only on the inter-particle distances. Numerical integrations over the parameters $\beta_i$ are required to complete the calculation of $V_C$ and $V_Q$.